\documentclass[12pt]{article}

\usepackage{amsmath,epsfig,epsf,psfrag,natbib}
\usepackage{amssymb}
\usepackage[latin1]{inputenc}

\usepackage{amsbsy}
\usepackage{lscape, epstopdf}
\usepackage{color,latexsym,amsfonts,amssymb,amsthm,amscd,amsmath,graphicx}

\newcommand{\be}{\begin{equation}}
\newcommand{\ee}{\end{equation}}
\newtheorem{theorem}{Theorem}
\newtheorem{lemma}{Lemma}
\newtheorem{example}{Example}

\newtheorem{proposition}{Proposition}

\newtheorem{corollary}{Corollary}
\newcommand{\bx}{\mbox{\bf x}}

\renewcommand{\theequation} {\arabic{section}.\arabic{equation}}

\newcommand{\ba}{\begin{eqnarray}}
\newcommand{\ea}{\end{eqnarray}}
\newcommand{\bas}{\begin{eqnarray*}}
\newcommand{\eas}{\end{eqnarray*}}

\newcommand{\e}{ { \mathbb{E}}}
\newcommand{\var}{ {\mathbb{V}\rm ar }}
\newcommand{\cov}{ {\mathbb{C}\rm ov}}
\newcommand{\bfe}{\mbox{\bf e}}

\newcommand{\bX}{\mbox{\bf X}}
\newcommand{\ben}{\begin{enumerate}}
\newcommand{\een}{\end{enumerate}}
\newcommand{\pr}{ \mbox{\rm pr}}
\def\T{{ \mathrm{\scriptscriptstyle \top} }}

\begin{document}
\date{}
\title{Full-semiparametric-likelihood-based inference for non-ignorable missing data}
\vspace{-0.1in}

\author{Yukun Liu, Pengfei Li, and Jing Qin\footnote{
Yukun Liu is  Professor,
School of Statistics,
       East China Normal University,
       500 Dongchuan Road,
       Shanghai 200241, China
(Email: \emph{ykliu@sfs.ecnu.edu.cn}).
Pengfei Li is Professor,
Department of Statistics and Actuarial Sciences,
University of Waterloo,
Waterloo, ON, Canada, N2L 3G1
(Email: \emph{pengfei.li@uwaterloo.ca}).
Jing Qin is Mathematical Statistician,
National Institute of Allergy and Infectious Diseases,
National Institutes of Health, 6700B Rockledge Drive MSC 7609,
Bethesda, MD 20892 (Email: \emph{jingqin@niaid.nih.gov}).
}
}

\maketitle

\vspace{-0.5in}
\begin{abstract}
During the past few decades, missing-data problems have been studied extensively,
with a focus on the ignorable missing case, where the missing probability
depends only on observable quantities. By contrast, research into
 non-ignorable missing data problems is quite limited. The main difficulty
in solving such problems is that the missing probability and
the regression likelihood function are tangled together in the
likelihood presentation, and the model parameters may not be
identifiable even under strong parametric model assumptions.
In this paper we discuss a semiparametric model for non-ignorable
missing data and  propose a maximum full semiparametric likelihood
estimation method, which is an efficient combination of the parametric
conditional likelihood and the marginal nonparametric biased sampling
likelihood. The extra marginal likelihood contribution can not only
produce  efficiency gain but also identify the underlying model
parameters without additional assumptions. We further show that
the proposed estimators for the underlying parameters and the
response mean are semiparametrically efficient. Extensive simulations
and a real data analysis demonstrate the advantage of the proposed method over competing methods.
\end{abstract}


%
\section{Introduction}
Missing data is a ubiquitous problem in many areas, such as survey sampling, epidemiology,
economics, sociology, and political science. Data may be missing because, for example,
not every individual is sampled due to cost or inconvenience,
or a sampled individual fails to report critical statistics.
Missing-data problems have been studied extensively during the last few decades.
Most research focuses on missing data that are ignorable or missing
at random in the sense that the missing probability or propensity score is
a function only of the observed data \citep{Little1987, Little2002, Rubin1987}.

Non-ignorable missing  or missing-not-at-random data occur if  the propensity score depends
on the missing data, even conditionally on the observed data.
Let $D$ be the missing indicator of  the variable of interest $Y$
associated with some covariate variables $\bX$, and $D=1$ if $Y$ is observed  and $D=0$ otherwise.
Non-ignorable missing implies that the propensity score $\pr(D=1|\bx,y)=\pr(D=1|\bX=\bx,Y=y)$
depends on  $y$ and possibly on $\bx$.
Inference for non-ignorable missing data is more challenging than
that for ignorable missing data for at least two reasons.
First,  the equality $\pr(y|\bx, D=1) = \pr(y|\bx, D=0)$, which holds
for ignorable missing data,   does not hold for non-ignorable missing data.
This implies that simply ignoring the missing data can lead to substantial selection bias
\citep{Groves2004}. Second,   unlike the ignorable missing case,
the propensity score and the regression likelihood function
are tangled together in non-ignorable missing-data problems,
and hence cannot be estimated separately.

These challenges require new modelling strategies for non-ignorable missing data.
The most popular strategy is to  make assumptions about  $\pr(D=1|\bx,y)$  and $\pr( y|\bx)$,
  based on the selection model factorization
$\pr(y, D|\bx)   = \pr(D|\bx,y) \pr( y|\bx)$  of \cite{Little1987, Little2002}.
Postulating parametric models \citep{Greenlees1982, Baker1988, Liu2010}
on both $\pr(D=1|\bx,y)$  and $\pr( y|\bx)$
is at risk of model mis-specification \citep{Little1985}.
Meanwhile completely nonparametric models for  $\pr(D=1|\bx,y)$
and $\pr(y|\bx)$ may  make  some model parameters not  identifiable \citep{Robins1997}.
Attention has been paid to the case where one of these probabilities
is parametric or semiparametric and the other is left unspecified.
\cite{Tang2003} made a parametric assumption
for $\pr(y|\bx)$  and left $\pr(D=1|\bx,y)$  unspecified.
Many researchers considered  a parametric model for $\pr(D=1|\bx,y)$
and a nonparametric model for $\pr(y|\bx)$.
See \cite{Qin2002, Chang2008, Kott2010, Morikawa2016, Morikawa2017} and \cite{Ai2018}.
\cite{Kim2011} proposed linking $\pr(y|\bx, D =1)$ and  $\pr(y|\bx, D =0)$
by a semiparametric exponential tilting model,
but a validation sample is required to estimate
the tilting parameter in the semiparametric model.
An alternative approach is to make parametric model assumptions
on the observed $Y$ given $\bX$ \citep{Lee2000, Riddles2016}.
An obvious advantage of this model over  a completely parametric model for $\pr(y|\bx)$
is that it  is checkable with available data.

Another layer of complication in non-ignorable missing-data problems is that
the underlying parameters for the propensity score and regression likelihood
function may not be identifiable even  if both of them are assumed
to follow completely parametric models.
\cite{Wang2014} found that
the model parameters can be identifiable given an instrument variable,
which is correlated with  the response variable but
is independent of the propensity score conditional on the response variable.
This strategy was further used by
\cite{Zhao2015} to identify  the model parameters
in generalized linear models based on non-ignorable missing data.
\cite{Miao2016b} systematically investigated
the identification of non-ignorable missing data via an ancillary variable,
which is equivalent to \cite{Wang2014}'s instrument variable.
Their general finding is that under non-ignorable missing data,
many commonly used models are identifiable,
and thus lack of identification is not an issue in many situations.

There have been many estimation approaches for identifiable model parameters
developed in recent years,  including
pseudo-likelihood approaches \citep{Tang2003, Zhao2015},
empirical likelihood method \citep{Zhao2013, Tang2014},
and the generalized method of moments with an instrument variable
\citep{Wang2014, Shao2016, Shao2018}. See \cite{Tang2018}  for a review  
of the most recent advances  in dealing with nonignorable missing data. 
Under parametric models for both the propensity score and
the observed $Y$ given $\bX$, \cite{Riddles2016} proposed
an  estimating equation method based on \cite{Louis1982}'s mean score equation,
although their estimator is generally not efficient.
When assuming a parametric model only for $\pr(D=1|\bx,y)$,
\cite{Morikawa2017} proposed an estimating equation method
by plugging-in a nonparametric estimator of $\pr(y|\bx, D=1)$,
which suffers from the curse of dimensionality and requires a bandwidth selection.
\cite{Morikawa2016} derived the semiparametric efficiency lower
bound under the same assumptions.
However, their semiparametrically efficient estimator requires explicit non-parametric estimation,
 also suffering from the curse of dimensionality and requiring a bandwidth selection.
 To avoid this dilemma,  \cite{Ai2018} proposed a new estimation method based
on the generalized method of moments with a diverging number of estimating equations.
As the number of estimating equation increases,
 their estimator attains the semiparametric efficiency lower bound
of  \cite{Morikawa2016}.
However, the constrained
generalized method of moments may have numerical convergence problems,
especially when some of the estimating equations are highly correlated.

In this paper, we consider parametric models for both $\pr(y|\bx, D=1)$ and  $\pr(D=1|\bx,y)$.
In particular, we assume that   $\pr(D=1|\bx,y)$  follows a logistic  regression model,
\begin{equation}
\label{logistic}
\pr(D=0|\bx,y)=
\frac{\exp(\alpha^*+\bx^\T \beta+y \gamma)}{
1+\exp(\alpha^*+\bx^\T\beta+y \gamma)},
\end{equation}
which is  commonly used in practice.
Under these assumptions, we find that the two distribution pairs
$\{ \pr(y|\bx, D=1), \; \pr(y|\bx, D=0)\}$  and $\{\pr(\bx|D=1), \;  \pr(\bx|D=0) \}$
satisfy two  density ratio models
\citep[DRMs]{Anderson1979},
see Equations \eqref{drm1} and  \eqref{drm2},
which share some key unknown parameters.
We give an easy-to-check condition to verify the identifiability of  the model parameters.
This condition is satisfied by many existing identification conditions
such as the   existence  of an instrument or ancillary variable
\citep{Wang2014, Miao2016a}.
For parameter estimation, the completely observed covariate data  can be used
to estimate the key unknown parameters, which
can be further used to estimate  $ \pr(y|\bx, D=0)$,
since $\pr(y|\bx, D=1)$ can be estimated directly using
the conditional maximum likelihood method.
These, together with the empirical distribution of $D$,
lead to estimation of  the conditional density $\pr(y|\bx)$;
consequently the characteristics of $Y$ can be consistently estimated.

Given the  completely observed covariate data and the fact that
$\{\pr(\bx|D=1), \; \pr(\bx|D=0) \}$ follows a DRM,
we use  \cite{Owen1988, Owen2001}'s empirical likelihood (EL)
to estimate the underlying parameters.
Since Owen's seminal paper,  the EL has become remarkably popular because
it has many nice   properties corresponding to those of
parametric likelihood methods, e.g., it is range-preserving, transformation-respecting,
and Bartlett correctable and it obeys Wilks' theorem
\citep{Hall1990, DiCiccio1991, Qin1994}.
The DRM-based  EL has been demonstrated to be very flexible and efficient,
and it has attracted much attention in recent decades;
see  \cite{Qin1997}, \cite{Chen2013}, \cite{Cai2017},  and the references therein.

We show that the maximum EL estimators of the underlying parameters
are asymptotically normal, and  the EL ratio for all the parameters follows
an asymptotically central chisquare distribution.
This makes it much more convenient to conduct hypothesis testing or {construct}
confidence intervals for these parameters.
We propose a maximum likelihood estimator (MLE)  for the marginal mean of
the response variable, and we establish  its asymptotic normality.
  We further show that the proposed MLEs for all parameters
attain the corresponding semiparametric efficiency lower bounds
under parametric assumptions for the propensity score
and the conditional density of $Y$ given $\bX$ and $D=1$.
Compared with the existing methods, the proposed maximum semiparametric
full likelihood approach has at least the following advantages:
\ben
\item
It is able to identify the underlying parameters whether
an instrument variable exists or not if the conditions in Proposition \ref{identifiability}
are satisfied.
The methods of \cite{Shao2016},
\cite{Riddles2016}, \cite{Morikawa2017},
\cite{Morikawa2016} and \cite{Ai2018}
 all require an instrument variable.
Further,  it is able to produce consistent estimators
for all the model parameters, if they are identifiable.
Extra information about the parameter $\gamma$ in (\ref{logistic}) is not needed.

\item
It applies to data of any dimension and is free of bandwidth selection.
The methods of \cite{Kim2011}, \cite{Shao2016}, \cite{Morikawa2016},
and \cite{Morikawa2017} all suffer from
the curse of dimensionality and  bandwidth selection,
  and may not work well for multivariate covariates.
\cite{Ai2018}'s method has an increasing calculation burden
as the number of estimating equation increases.

\item
Existing methods handling non-ignorable missing-data problems under semiparametric
setups are mainly based on estimating equations and may not be the most efficient in general.
Since full likelihood approaches are generally the most efficient,  it can be expected that
the proposed maximum semiparametric full likelihood approach would outperform the existing methods.
Even though \cite{Morikawa2016} calculated the semiparametric efficiency lower bound
with the specification of propensity score only, their lower bound is not achievable
unless the  conditional density of $Y$ given $(\bX, D=1)$ is fully specified.
In this paper we show that with the knowledge of $\pr(y|\bx, D=1)$,
\cite{Morikawa2016}'s method is no longer optimal anymore.
Our new lower bound is lower than theirs.

\item
Our method  is also applicable to retrospectively collected data.
For example,  when the number of nonresponse individuals (with $D=0$) is large,
we can randomly select some covariate $\bx$ from them to save cost.
Based on  this data together with the fully observed data,
our method  still provides valid inference about the underlying population.
However, the existing methods may produce biased estimators
because they are designed for prospective data.
\een

The rest of this paper is organized as follows.
In Section 2, we introduce the proposed model,
 show its equivalence  to  two DRMs,  and provide sufficient conditions for
the identifiability of the model parameters.
Section 3 presents the proposed  semiparametric DRM-based
EL  method and   the resulting MLEs for the underlying parameters and
the  mean of the  response variable;
Their asymptotic normalities   and semiparametric efficiencies are    also established.
Section 4 reports extensive simulation results.
A real-life set  of data is  analyzed for illustration in Section 5.
  Section 6 provides concluding remarks.
All  technical details are given in the Appendix.

\section{Model and its identifiability}

\subsection{Model set-up}
Suppose $\{ (y_i, \bx_i, d_i), i=1, \ldots, n\}$ are $n$ independent and identically distributed
copies of $(Y, \bX, D)$, where the covariates $\bx_i$  are always observed,
and $y_i$ is observed if and only if $d_i=1$.
We assume that the missing probability satisfies the logistic regression model in (\ref{logistic}), i.e.,
\begin{equation*}
\pr(D=0|\bx,y)=\frac{\exp(\alpha^*+\bx^\T \beta+y \gamma)}{
1+\exp(\alpha^*+\bx^\T\beta+y \gamma)}.
\end{equation*}
 The parameter  $\gamma$ is called the tilting  parameter \citep{Kim2011}.
 It quantifies the extent to which the model departs from  ignorable missing,
and  $\gamma=0$  corresponds to the ignorable missing-data case.
We are interested in estimating the underlying parameters $(\alpha^*, \beta, \gamma)$
and the marginal mean $\mu$ of $Y$.

Based on the observed data, the full likelihood is
\begin{equation}
\label{full.lik}
 \prod_{i=1}^n\left[ \{\pr(D=1|\bx_i,y_i)\pr(y_i,\bx_i) \}^{d_i}
\left\{ \int \pr(D=0|\bx_i,y)\pr(y,\bx_i)dy  \right\}^{1-d_i}\right].
\end{equation}
Unlike the case of ignorable missing,
here $\pr(D=1|\bx,y)$ and $\pr(y, \bx)$ can not be separated
and hence can not be separately estimated.
To make inference based on the full likelihood,
{one may postulate} parametric assumptions on $\pr(D=1|y,\bx)$ and $\pr(y|\bx)$, which
are sensitive to model mis-specification
\citep{Little1985, Kenward1988}.

We crack this nut by an alternative method.
The logistic regression model \eqref{logistic}
is equivalent to the two-sample DRM \citep{Qin1997}
\ba
\label{drm-xy}
\pr(\bx,y|D=0)=\exp(\alpha + \bx^\T \beta+y\gamma)   \pr(\bx,y|D=1),
\ea
where $\alpha  = \alpha^* + \log\{\eta/(1-\eta) \}$
and $\eta =\pr(D=1)$ is the probability of being observed.
Clearly,  $\eta$ can be consistently estimated by data and is therefore identifiable.
Then the identifiability of $\alpha^*$ is equivalent to that of $\alpha$.

Integrating out $y$, we have
\begin{eqnarray*}
\pr(\bx|D=0) 
&=&\exp(\alpha+\bx^\T \beta)\pr (\bx|D=1) \int \exp(y\gamma)\pr(y|\bx,D=1)dy.
\end{eqnarray*}
Therefore, the conditional densities of $Y=y$ given $(\bX=\bx, D=0)$ and given $(\bX=\bx,D=1)$ satisfy
\ba
\label{drm-star}
\pr(y|\bx, D=0)
= \frac{\pr(\bx,y|D=0)}{\pr(\bx|D=0)}
=\frac{\exp(y\gamma)\pr(y|\bx, D=1)}{\int \exp(y\gamma)\pr(y|\bx, D=1)dy}.
\ea
Although $\pr(y|\bx,D=1)$ is directly estimable based on the observed
$(y_i, \bx_i)$'s with $d_i=1$, it is impossible to estimate $\pr(y|\bx,D=0)$
since $\gamma$ is unknown in general.
As a consequence, the conditional approach
is not viable, as demonstrated by \cite{Kim2011}, who rely on
external data to identify $\gamma$.  In practical applications,
however, external data are often unavailable, which makes the estimation
of $\gamma$ impossible.

Fortunately, the marginal information
on the $(\bx_i,d_i)$'s can help to identify $\gamma$,
which  solves the thorny identifiability problem in non-ignorable missing-data problems.
Since $(y_i, \bx_i)$'s with $d_i=1$ are available,
without loss of generality, we can postulate a parametric
model $f(y|\bx, \xi)$ for $\pr(y|\bx, D=1)$ with an identifiable parameter $\xi$.
The parameter $\xi$ can be  consistently estimated
from the directly observed data.
This parametric model  together with  Equation~\eqref{drm-star} implies two DRMs:
\ba
\label{drm1}
\pr(y|\bx,D=0)&=&
 \exp\{ \gamma y - c(\bx,\gamma,\xi)  \} f (y|\bx,\xi),\\
\label{drm2}
\pr(\bx|D=0)
&=& \exp\{ \alpha + \bx^\T\beta+c(\bx,\gamma,\xi) \} \pr(\bx|D=1),
\ea
where
\ba
\label{c-fun}
 c(\bx,\gamma,\xi)  =  \ln \left\{  \int \exp(y\gamma) f (y|\bx,\xi) dy \right\}.
\ea
Equations \eqref{drm1}--\eqref{c-fun} are
the foundation of our inference method. We note that the second DRM
involves all the underlying parameters in the model
and is dependent only on  $\pr(\bx|D=0)$ and   $\pr(\bx|D=1)$.
Since the $(\bx_i, d_i)$'s  with $d_i=0$ or 1 are not subject to missingness,
the parameters can be consistently estimated by  their
maximum DRM-based EL estimators \citep{Qin1997}  provided they are identifiable.

 \subsection{Model identifiability}

\cite{Miao2016a} pointed out that
even under full parametric models for $\pr(D=1|\bx, y)$ and $\pr(y|\bx)$,
the underlying model parameters may not be identifiable.
This phenomenon also arises under Model \eqref{drm2},
where even $\pr(\bx|D=1)$ is completely known, the model parameters in \eqref{drm2}
may not be identifiable.
We present a simple-to-check sufficient condition  for
the identifiability of the underlying parameters in \eqref{drm2}.
We have {assumed} that   $ \xi$ is identifiable.  Hence, we focus here on
the identifiability of   the parameters $\alpha, \beta$, and $\gamma$.
Given the data
 $ \{ (\bx_i,d_i),  i=1, \ldots, n \}$,
the conditional density functions   $\pr(\bx|D=0)$  and  $\pr(\bx|D=1)$
are clearly identifiable  and can be consistently estimated by, for example,
the kernel method.
The log ratio  $\log\{ \pr(\bx|D=0)/\pr(\bx|D=1) \}$ is also identifiable.
Since
\[
 \log\{ \pr(\bx|D=0)/\pr(\bx|D=1) \} =  \alpha + \bx^\T\beta+c(\bx,\gamma,\xi),
\]
 the model identification is equivalent to
the identification of the parameters $\alpha, \beta$, and $\gamma$
in $\alpha + \bx^\T\beta+c(\bx,\gamma,\xi)$.

\begin{proposition}
\label{identifiability}
Let $S$ be  the common support of $\pr(\bx|D=0) $ and $\pr(\bx|D=1)$, and
\(
\Omega = \{ h(\bx):  S  \mapsto \mathbb{R}\mid \exists
 (\alpha, \beta,   \gamma) \ \mbox{such that }
h(\bx) =  \alpha + \bx^\T\beta+c(\bx,\gamma,\xi) \ \forall \ \bx \in S
  \}.
\)
If for any  $h(\bx)\in \Omega$, there exists a unique $ (\alpha, \beta,   \gamma)$
such that $h(\bx) =  \alpha + \bx^\T\beta+c(\bx,\gamma,\xi)$,
then  $(\alpha,\beta,\gamma)$ is identifiable.
\end{proposition}

Next we apply the above proposition to some special cases.
We need the concept of an instrument variable, which can be helpful to identify $\gamma$.
Suppose $\bx$ can be written as  $\bx=(z, u^\T)^\T$. If
\bas
\pr(D=0| z, u, y)
&=&
\pr(D=0| u, y)
=
\frac{\exp(\alpha^*+ u^\T \beta +y \gamma)}{1+\exp(\alpha^*+ u^\T\beta+y \gamma)}
\eas
and  $\pr(y|\bx)=\pr(y|z, u)$ depends on $z$ and possibly on $u$, then $z$ is an instrument variable.
That is, an instrument variable is defined to be a covariate that
 does not affect  the missingness but may affect the conditional distribution of the response variable.

 With the above preparation and Proposition \ref{identifiability},
 we find that $(\alpha,\beta,\gamma)$ is identifiable in the following two cases.

\begin{corollary}
\label{identifiability2}
Suppose the logistic regression model in
\eqref{logistic} holds
and that the density function of
 $Y$ given $(\bX=\bx, D=1)$ is $f(y| \bx, \xi)$.
 (a)  If there exists {an instrument variable $z$} in $\bx$,
 then $(\alpha,\beta,\gamma)$ is identifiable.
 (b)    Assume that  the set $S$  in Proposition \ref{identifiability}
 contains an open set, and $c(\bx,\gamma,\xi)$ can be expressed as
$ c(\bx,\gamma,\xi)= \sum_{i=1}^{k}  a_{i}(\gamma) g_i(\bx)
+ a_{k+1}(\gamma) + \bx^\T a_{k+2}(\gamma) $
for some positive integer $k$, and  continuous functions
$a_i(\gamma)$ ($i=1,\ldots,k+2$) and $g_i(\bx)$ ($i=1,\ldots,k$),
where  $1, \bx, g_1(\bx), \ldots, g_{k}(\bx)$  are linearly independent, and
$a_{j}(\gamma)$ ($j=1, \ldots, k$) are not equal to the zero functions.
If $\big(a_1(\gamma_1), \ldots, a_{k}(\gamma_1)\big) \neq \big(a_1(\gamma_2),
 \ldots, a_{k}(\gamma_2)\big)$ for any $\gamma_1\neq \gamma_2$,
 then $(\alpha,\beta,\gamma)$ is identifiable.

 \end{corollary}

 As an application of the above results, we consider the normal model
 in which  $f(y|\bx, \xi)$ is  the density function of
$N\big(\mu(\bx,\xi), \sigma^2(\bx,\xi)\big)$.
 Direct calculations give
$
c(\bx, \gamma, \xi)
=    \gamma   \mu(\bx,\xi) +  0.5 \gamma^2 \sigma^2(\bx,\xi).
$
 Further, assume    $\mu(\bx,\xi)=\bx^\T b_1(\xi)+ b_2(\xi)  \bx^\T \bx  $
and  $\sigma^2(\bx,\xi)=   \exp\{b_3(\xi)+ \bx^\T b_4(\xi) \}$
for  nonzero  functions $b_i(\xi)$.
We have the following observations:
\begin{description}
\item[(I)]  If  $b_2(\xi) \neq 0$, then
according to Corollary \ref{identifiability2},
$(\alpha, \beta, \gamma)$ is identifiable.
\item[(II)] If   $b_2(\xi ) = 0$  and $b_4(\xi)=0$, then
$$
 {\alpha + \bx^\T\beta+c(\bx,\gamma,\xi)}
=\alpha+0.5 \gamma^2 \exp\{b_3(\xi)\}+ \bx^\T\{\beta+\gamma b_1(\xi)\},
$$
which together with Lemma \ref{identifiability} implies that $(\alpha, \beta, \gamma)$ is not identifiable.
\item [(III)] If   $b_2(\xi ) = 0$  and $b_4(\xi)\neq0$, then
$$
{\alpha + \bx^\T\beta+c(\bx,\gamma,\xi)}
=\alpha+\bx^\T\{\beta+\gamma b_1(\xi)\}+0.5 \gamma^2 \exp\{b_3(\xi)+\bx^\T b_4(\xi) \}.
$$
If further $\gamma=0$, then Proposition \ref{identifiability} implies
that $(\alpha, \beta, \gamma)$ is  identifiable. Otherwise,  $(\alpha, \beta, \gamma)$ is not identifiable.
\end{description}

\section{Semiparametric empirical likelihood inference}

\subsection{ Empirical likelihood}

Suppose there are $n_1$ completely observed data and $n_2$ partially observed data.
Without loss of generality, we assume that $d_i=1$, $i=1,\ldots,n_1$
and $d_i=0$, $i=n_1+1,\ldots,n$.
The full likelihood in (\ref{full.lik}) can be  written as
\bas
\prod_{i=1}^{n_1}  \{  \pr(y_i|\bx_i, D=1)\pr(\bx_i|D=1)\pr(D=1)  \} \cdot
\prod_{i=n_1+1}^{n}  \{\pr(\bx_i|D=0)\pr(D=0)\}.
\eas
Let $\theta = (\alpha, \beta^\T, \gamma, \xi^\T)^\T$ and
$t(\bx, \theta) = \alpha +\bx^\T\beta+c(\bx,\gamma,\xi)$.
Since   $\pr(y |\bx,D=1 ) = f(y| \bx,\xi)$ by assumption,
it follows from $\eta =\pr(D=1)$ and Equation \eqref{drm2} that
the full log-likelihood is
$
\tilde \ell
= \ell_1 (\eta) +\tilde \ell_2,
$
where
\[
\ell_1 (\eta) =  n_1\log(\eta) + (n-n_1) \log (1-\eta)
\]
is the marginal likelihood based on the $d_i$'s,
and
\bas
\tilde \ell_2
 &=& \sum_{i=1}^{n_1}\log \{ f(y_i|\bx_i, \xi) \}
  + \sum_{i=n_1+1}^n  t(\bx_i, \theta) +\sum_{i=1}^n\log \{ \pr(\bx_i|D=1) \}
\eas
is a conditional likelihood given the $d_i$'s.

We leave the conditional density  $\pr(\bx|D=1)$ completely unspecified,
and use the celebrated EL method of \cite{Owen1988, Owen1990} to handle it.
Let $p_i=\pr(\bx_i|D =1)=dF(\bx_i|D =1) $, where $F(\bx |D =1) $
is the cumulative distribution function
corresponding to the density $\pr(\bx|D =1)$.
Following the principle of EL,  $\tilde \ell_2$ becomes an empirical log-likelihood
\bas
\tilde  \ell_2
= \sum_{i=1}^{n_1}\log \{ f(y_i|\bx_i, \xi) \}
  + \sum_{i=n_1+1}^n  t(\bx_i, \theta) +\sum_{i=1}^n\log( p_i),
\eas
where the $p_i$'s are subject to the constraints
\[
p_i\geq 0, \quad
\sum_{i=1}^np_i=1, \quad
\sum_{i=1}^np_i[\exp\{ t(\bx_i, \theta) \}-1]=0.
\]

Maximizing $\tilde \ell_2 $ with respect to the $p_i$'s,
we arrive at
\begin{equation}
\label{mle.phat}
p_i =\frac{1}{n} \frac{1}{1+ \lambda[\exp\{  t(\bx_i, \theta) \}-1]},
\end{equation}
where $\lambda$ is the solution to
\begin{equation}
\label{mle.lambda}
\sum_{i=1}^n  \frac{\exp\{  t(\bx_i, \theta) \}-1}{1+ \lambda[\exp\{ t(\bx_i, \theta)\}-1]} = 0.
\end{equation}
Substituting these $p_i$'s into $\tilde \ell_2$ leads to the profile log-likelihood of $\theta$,
\bas
\ell_2(\theta)
=
  \sum_{i=1}^{n_1}\log \{  f(y_i|\bx_i, \xi)\} + \sum_{i=n_1+1}^n  t(\bx_i, \theta)
- \sum_{i=1}^{n} \log\big\{ 1+ \lambda [\exp\{ t(\bx_i, \theta)\}-1]\big\}.
\eas
The profile log-likelihood of $(\eta, \theta)$ is then defined as
\begin{equation}
\label{profile.lik}
\ell(\eta, \theta)=\ell_1(\eta)+\ell_2(\theta).
\end{equation}

\subsection{Estimation of the underlying parameters}
With the profile log-likelihood of $(\eta, \theta)$  in (\ref{profile.lik}),
the MLE of $(\eta, \theta)$ is
$$
 (\hat\eta, \hat \theta)
 =\arg\max_{\eta, \theta}\ell(\eta, \theta).
$$
Equivalently, $\hat\eta$ maximizes $\ell_1(\eta)$, which gives $\hat \eta= n_1/n$,
and
$
 \hat \theta  = (\hat \alpha, \hat \beta^\T, \hat \gamma, \hat \xi^\T)^\T
=  \arg\max_{\theta} \ell_2(\theta ).
$
The likelihood ratio function of $ \theta$
is  defined  as
\bas
R( \theta)
= 2 \{\max_{\eta,  \theta}\ell(\eta,  \theta)- \max_{\eta}\ell(\eta,  \theta) \}
=2\{  \ell_2(\hat \theta) -  \ell_2( \theta )  \}.
\eas

Next we  study the large-sample properties of the MLE and the likelihood ratio.
Denote the truth of $(\eta, \theta)$
by $(\theta_0, \eta_0)$ with $\theta_0=(\alpha_0, \beta_0^\T, \gamma_0, \xi_0^\T)^\T$
and $\eta_0 \in (0, 1)$.
Define
\bas
\pi(\bx; \theta, \eta)
&=& \frac{ (1-\eta)  \exp\{t(\bx,\theta)\}}{
    \eta + (1-\eta) \exp\{ t(\bx,\theta) \} }
\eas
and we write $\pi(\bx)=\pi(\bx; \theta_0, \eta_0)$ for abbreviation.
  Let  $d_{\theta}$  denote the  dimension of  $\theta$ and
${\bf e}_1$ be a $d_{\theta}\times 1$ vector with the first component being 1
and the remaining components 0.
Finally, define
\ba
\label{var.mat}
V =   \e[  \{1-\pi(\bX)\} \pi(\bX)   \{ \nabla_\theta t(\bX, \theta_0  ) \}^{\otimes 2} ]
+
 \e[  D  I_e \{ \nabla_{\xi } f(Y|\bX,  \xi_0 ) \}^{\otimes 2}I_e^\T ],
\ea
where $\nabla_{\theta}$ is the differentiation operator with respect to $\theta$,
\(
I_e^\T  = (0_{d_{\xi}\times (2+d_{\beta})}, I_{d_\xi\times d_{\xi}})
\),
and $B^{\otimes 2} = BB^\T$ for any matrix or vector $B$.

\begin{theorem}
\label{asy-par}
Assume Conditions A1--A4 in Appendix 1.
Suppose that the logistic regression model in
\eqref{logistic} holds with $(\alpha_0, \beta_0, \gamma_0)$
in place of $(\alpha, \beta, \gamma)$,
and that the density function of
 $Y$ given $(\bX=\bx, D=1)$ is $f(y| \bx, \xi_0)$.
 Further, assume that $\theta$ is identifiable.
Then as $n\rightarrow \infty$,
(1)
$\sqrt{n} (\hat \theta -\theta_0)
\to N\big(0, \; V^{-1}- \{\eta_0(1-\eta_0)\}^{-1} {\bf e}_1{\bf e}_1^\T  \big) $ in distribution with $V$
defined in (\ref{var.mat});
(2) $ R(\theta_0 )  \to \chi_{d_{\theta}}^2$ in distribution.

\end{theorem}

Theorem \ref{asy-par} implies that the MLEs of all the parameters are  asymptotically normal.
The likelihood ratio for the parameters  follows a central chisquare limiting distribution,
which makes the resulting hypothesis testing  or interval estimation
about $ \theta $ very convenient.
Although the proposed approach is developed based on prospective data,
we emphasize that  it can also apply to retrospectively collected data.
This is because the subsequent inferences are mainly based on
 $  \ell_2$ or equivalently
\bas
\tilde \ell_2
&=&
\log\left[\prod_{i=1}^{n_1}  \{  \pr(y_i, \bx_i|D=1)   \}
\prod_{i=n_1+1}^{n}  \{\pr(\bx_i|D=0)  \right],
\eas
which  is actually a retrospective log-likelihood.
If   $\eta=\pr(D=1)$ or $\hat \eta$ is available,
based on  retrospectively collected data,
the proposed approach can still make valid inference.

Given the MLE of all the underlying parameters,
we are able to construct the MLE of the population mean  $\mu$ of the response $Y$.
Under our model,   $\mu$ depends not only on
the underlying parameters $ \theta $  but also on
$\pr(\bx|D=1)$ or the corresponding cumulative distribution function
$F(\bx|D=1)$. With  the MLEs $ \hat  \theta$ and $\hat \eta=n_1/n$,
we show in the supplementary material that $\hat\lambda=n_2/n$,
where $\hat\lambda$ satisfies (\ref{mle.lambda}) with $\hat  \theta $
in the place of $ \theta$.
With (\ref{mle.phat}), the MLE of $p_i$ is
\bas
\hat p_i
= \frac{1}{n} \frac{1}{1+ (n_2/n)[\exp\{t(\bx_i,  \hat \theta)\}-1]}
= \frac{1}{n} \frac{1}{\hat \eta+ (1-\hat \eta)
 \exp\{ t(\bx_i,  \hat \theta)\} }.
\eas
Accordingly  the MLE of $F(\bx|D=1)$ is
$
\hat F(\bx|D=1) = \sum_{i=1}^{n_1} \hat p_i I(\bx_i\leq \bx),
$
where for two vectors $\bx_1$ and $\bx_2$,  $\bx_1\leq \bx_2$
implies that the inequality holds elementwise.

\subsection{Estimation of the response mean}

To obtain the MLE of  the response mean $\mu$,
we write $\mu$ in terms of  the underlying parameters   $\eta,
\theta$, and $F(\bx|D=1)$ as follows:
\bas
\mu
&=&
 \int_y \int_{\bx} y \pr(y|\bx, D=1) \pr(\bx|D=1) \pr(D=1) d\bx dy \\
&& +
 \int_y \int_{\bx} y \pr(y|\bx, D=0) \pr(\bx|D=0) \pr(D=0) d\bx dy
\\
&=&
 \int_y \int_{\bx} y \pr(y|\bx, D=1) \pr(\bx|D=1) \eta d\bx dy \\
 && +
\int_y \int_{\bx} y  \exp(\alpha + \bx^\T\beta  +\gamma y)  \pr(y|\bx, D=1) \pr(\bx|D=1) (1-\eta) d\bx dy
\\
&=&
\int_{\bx} \left[ \int_y  y \{ \eta +(1-\eta)\exp(\alpha + \bx^\T\beta  +\gamma y) \} f(y|\bx, \xi)dy\right] dF(\bx|D=1),~~~~~
\eas
where in the last step we replace $\pr(y|\bx, D=1) $
and $\pr(\bx|D=1) d\bx$ by
$f(y|\bx, \xi)$ and $dF(\bx|D=1)$, respectively.
Then the MLE of $\mu$ is
\ba
\hat \mu  &=&
\sum_{i=1}^n   \hat p_i
\left[ \int_y  y \{ \hat \eta +(1-\hat \eta)\exp(\hat \alpha
    + \bx_i^\T\hat \beta  +\hat \gamma y) \} f(y|\bx_i, \hat \xi)dy\right]  \nonumber \\
   &=&
  \frac{1}{n} \sum_{i=1}^n \frac{  \int_y  y \{ \hat \eta +(1-\hat \eta)
       \exp(\hat \alpha + \bx_i^\T\hat \beta  +\hat \gamma y) \} f(y|\bx_i, \hat \xi)dy  }{
  \hat \eta + (1-\hat \eta)  \exp\{ t(\bx_i,  \hat \theta)\} }.
  \label{mle1}
\ea

We use the normal model as an illustrating example:
$f(y|\bx, \xi)$ is chosen to be the density function of
$N\big(\mu(\bx,\xi), \sigma^2(\bx,\xi)\big)$.
In this example,  the proposed   mean estimator in (\ref{mle1}) becomes
\ba
\label{mu-normal}
\hat \mu
=
  \frac{1}{n} \sum_{i=1}^n \frac{ \hat \eta \hat\mu_i
  +(1-\hat \eta)
  ( \hat\mu_i+  \hat \gamma  \hat\sigma_i^2 )
  \exp(\hat \alpha + \bx_i^\T\hat \beta+\hat\mu_i\hat \gamma +  0.5  \hat \gamma^2\hat\sigma_i^2 )
     }{
  \hat \eta + (1-\hat \eta) \exp( \hat \alpha +\bx_i^\T\hat \beta+\hat\mu_i\hat \gamma +  0.5  \hat \gamma^2\hat\sigma_i^2) },
\ea
where   $\hat\mu_i= \mu(\bx_i, \hat\xi) $ and   $\hat\sigma_i^2 = \sigma^2(\bx_i, \hat\xi)$.


The next theorem establishes the asymptotic normality of the proposed estimator $\hat\mu$ in (\ref{mle1}).

\begin{theorem}
\label{asy-mu}
Under the conditions of Theorem \ref{asy-par}, as $n$ goes to infinity,
$\sqrt{n}(\hat\mu  - \mu) \to N(0, \sigma^2)$ in distribution,
where
\(
\sigma^2
= \var\{K(\bX;\theta_0,\eta_0) \}+  A^\T V^{-1} A
\)
with
\bas
K(\bx;\theta,\eta)
&=&
 \frac{
\int  y \{  \eta +(1-\eta) \exp(  \alpha + \bx^\T \beta  + \gamma y) \} f(y|\bx,   \xi)dy  }{
 \eta +(1-\eta)  \exp\{  \alpha +\bx^\T  \beta+c(\bx, \gamma,  \xi)\} }
\eas
and $A=\e\left\{ \nabla_{\theta}K(\bX;\theta_0,\eta_0)\right\}$.

\end{theorem}

When Wald-type intervals are constructed for $\mu$ based on Theorem \ref{asy-mu},
we need a consistent estimator of $\sigma^2$,
which can be constructed based on consistent estimators of
 $A$, $  \var\{K(\bX;\theta_0,\eta_0) \}$, and $V$.
Reasonable estimators for these three quantities are
\bas
\hat A &=& n^{-1}  \sum_{i=1}^n \nabla_{\theta}  K(\bx_i; \hat \theta,\hat \eta),\\
\widehat \var\{K(\bX;\theta_0,\eta_0) \}
&=&
n^{-1} \sum_{i=1}^n  \{ K(\bX; \hat \theta, \hat \eta) \}^2
-  \left\{n^{-1} \sum_{i=1}^n  K(\bX; \hat \theta, \hat \eta) \right\}^2,
\eas
and
$$
\hat V
=
n^{-1}  \sum_{i=1}^n  [ \{1-\pi(\bx_i, \hat \theta, \hat\eta)\} \pi(\bx_i, \hat \theta, \hat\eta)
 \{ \nabla_\theta t(\bx_i, \hat \theta  ) \}^{\otimes 2}
+
   d_i  I_e \{ \nabla_{\xi } f(y_i|\bx_i,  \hat \xi ) \}^{\otimes 2}I_e^\T  ].
$$
These estimators are consistent because $(\hat \theta, \hat \eta)$ is
consistent and $K$ is smooth in all its arguments.
Consequently a consistent estimator of $\sigma^2$ is
\ba
\label{sigma-hat}
\hat \sigma^2 =  \widehat \var\{K(\bX;\theta_0,\eta_0) \} + \hat A^\T \hat V^{-1} \hat A.
\ea

\subsection{Semiparametric efficiency}

  We  make the same model assumptions as \cite{Riddles2016}:
the logistic model in \eqref{logistic} for the propensity score
and  a parametric model $f(y|\bx, \xi)$ for $\pr(y|\bx, D=1)$,
and leave    $\pr( \bx|D=1 )$ completely unspecified.
 Therefore our model setup is  semi-parametric.
 Next we show that the estimators $(\hat \theta, \hat \eta)$ and $\hat \mu$,
 which are built on the above semi-parametric model,
are semiparametrically efficient.

\begin{theorem}
\label{semi-efficiency}
Under the conditions of Theorem \ref{asy-par},
the MLEs $(\hat \theta, \hat \eta)$ and $\hat \mu$
are both semiparametrically efficient in sense that
their asymptotic variances attain the corresponding
semiparametric efficiency lower bounds.
\end{theorem}

 We make some comments on Theorem \ref{semi-efficiency}
and the results in \cite{Riddles2016},  \cite{Morikawa2016} and \cite{Ai2018}.
Note that \cite{Riddles2016}'s estimator was constructed
under the same model assumptions as ours.
Theorem \ref{semi-efficiency} implies that
 the asymptotic variance of their mean estimator
is no less than $\sigma^2$,
the asymptotic variance of the MLE $\hat \mu$
and also the semiparametric
efficiency lower  bound for estimating $\mu$.

When only the parametric propensity score assumption is made,
\cite{Morikawa2016} derived   the
semiparametric  efficiency lower bound for the parameter of interest such as the response mean,
and proposed two adaptive estimators whose asymptotic variances attain the lower bound.
\cite{Ai2018} proposed an estimation method based
on the generalized method of moments,
and  showed that as the number of moments increases appropriately
their estimator also attains the lower bound of \cite{Morikawa2016}.
According to \cite{Tsiatis2006},  the semiparametric efficiency lower bound is
equal to the supremum of the asymptotic variances of the MLEs under all parametric submodels.
Since the model assumptions in \cite{Morikawa2016}
is weaker  than   ours,
the set of all parametric submodels considered in \cite{Morikawa2016}
contains all parametric submodels considered in this paper.
Consequently, when the parameter of interest is the response mean,
the semiparametric efficiency lower bound of \cite{Morikawa2016}
is no less than $\sigma^2$.
Hence the asymptotic variances of
\cite{Morikawa2016}'s two adaptive estimators and  \cite{Ai2018}'s estimator
 are no less than that of  our estimator $\hat \mu$.

\subsection{Model checking}

Based on the completely observed data $\{(y_i,\bx_i,d_i=1), i=1,\ldots,n_1\}$,
we can directly examine the correctness of the model assumption $\pr(y|\bx, D=1)=f(y|\bx,\xi)$.
One method for this is Ducharme and Ferrigno (2012)'s
omnibus  goodness-of-fit test  for conditional distributions.
Another question about the proposed model
is the reliability of the parametric model assumption
on the propensity score $\pr(D=1|\bx,y)$.
Since we do not observe $y_i$'s for $\{(\bx_i,d_i=0), i=n_1+1,\ldots,n\}$,
we do not have direct data to check this.
However, the question can be answered indirectly by
testing  the goodness-of-fit of the DRM \eqref{drm2}.
The latter problem has been studied by many researchers
and can be solved by the tests of
\cite{Qin1997}, \cite{Cheng2004}, \cite{Bondell2007},  and others.

\section{Simulation}

\subsection{Set-up}
We carry out simulations to investigate the finite-sample performance of
the proposed  estimator for   the population mean of the response.
We compare the proposed  mean estimator $\hat\mu$
with  four others: (1) \cite{Morikawa2016}'s adaptive estimator
with correctly specified parametric form for $\pr(y|\bx, D=1)$,   $\tilde \mu_t$,
(2) \cite{Morikawa2016}'s adaptive estimator
without specifying a parametric form for $\pr(y|\bx, D=1)$,   $\tilde \mu_{np}$,
(3) the sample mean of the observed response, $\bar y_r$,
and
(4) the sample mean of all the responses  when there are no missing data, $\bar y$.
When $\pr(y|\bx, D=1)$ is correctly specified,
\cite{Morikawa2016} showed that $\tilde \mu_t$ is more efficient than
\cite{Riddles2016}'s estimator,
and further \cite{Ai2018}'s estimator has the same asymptotic variance as $\tilde \mu_t$.
Hence \cite{Riddles2016}'s and \cite{Ai2018}'s methods are not included in the comparison.

We generate data from  the following three examples.
\begin{example}
\label{exm4.1}
Let $\bx=(z, u)^\T$, where $u$ is a Bernoulli random variable with  success probability 0.5,
$z$ follows the uniform distribution on $(-1,1)$, and $u$ and $z$ are independent.
We choose
$
\pr(D=1|\bx, y) = 1/\{1+\exp(-1.7-0.4u +0.5y)\}
$
and set
 $\pr(y|\bx, D=1) = f(y|\bx, \xi)$ to the density function of
$N\big(\mu(\bx), \sigma^2 \big)$, where
$
\mu(\bx) =\exp(0.5-u+1.5 z)
$
and
$
\sigma^2=1\mbox{ or }4.
$

\end{example}

\begin{example}
\label{exm4.2}
Let $\bx=(z, u)^\T$, where  $u\sim N(1,1)$, $z\sim N(0,1)$
and $u$ and $z$ are independent.
We choose
$
\pr(D=1|\bx, y) = 1/\{1+\exp(-1.7-0.4u +0.5y)\}
$
and set
 $\pr(y|\bx, D=1) = f(y|\bx, \xi)$ to the density function of
$N\big(\mu(\bx), \sigma^2 \big)$, where
$
\mu(\bx) =2.5-u+1.5 z
$
and
$
\sigma^2=1\mbox{ or }4.
$

\end{example}

\begin{example}
\label{exm4.3}
The covariate $x$  follows  $N(2, 1)$.
We choose
$
\pr(D=1|x, y) = 1/\{1+\exp(-2.7-0.4x+0.5 y)\}
$
and set $\pr(y|x, D=1) = f(y|x, \xi)$
to the density function of $N\big(\mu(x), \sigma^2 e^{0.5x} \big)$,
where
$
\mu(x) =  2- x+x^2
$
and
$
\sigma^2=1 \mbox{ or  } e^{0.7}.
$

\end{example}

Example \ref{exm4.1} is Scenario 2 of \cite{Morikawa2016}
except that we consider $\sigma^2=1$ and $4$, while \cite{Morikawa2016}
only considered  $\sigma^2=1$.
Example \ref{exm4.2} represents the case where the mean function
is a linear function  of $\bx$.
Both Examples \ref{exm4.1} and \ref{exm4.2} have an instrument variable
so the model parameters are  identifiable.
Example \ref{exm4.3} represents the case that there is no instrument variable,
but the model parameters are still identifiable.
The true values of $\mu$ and the missing probability $1-\eta$
for the three examples are tabulated in Table \ref{true.mu}.

\begin{table}[!ht]
\renewcommand{\arraystretch}{0.66}
\caption{
True values of $\mu$ and the missing probability $1-\eta$ in Examples \ref{exm4.1}--\ref{exm4.3}.
\label{true.mu}
}
\centering
{
\tabcolsep 4pt
\vspace{0.1in}
\begin{tabular}{cccc | cccc} \hline
Example &$\sigma^2$&$\mu$&$1-\eta$& Example &$\sigma^2$&$\mu$&$1-\eta$\\  \hline
1&1&1.748 &0.294 &  1&4&2.326&0.362\\
2&1&1.638 &0.275 & 2&4&2.177&0.339\\
3&1&3.129 &0.277 & 3&$e^{0.7}$&3.289&0.299\\  \hline
\end{tabular}
}
\end{table}

 \subsection{Point estimation}

In this section, we evaluate the performance of the five mean estimators in terms of
the relative bias (RB) and mean square error (MSE).
We set $n=500$ and $2000$ for all the three examples,
and use 2000 for the number of repetitions  in all our simulations.
The simulation results are summarized in Table \ref{RB.MSE}.

It is worth mentioning that we encountered some numerical problems
in the implementation of \cite{Morikawa2016}'s adaptive estimator $\tilde \mu_t$,
in Example \ref{exm4.1} with $n=500, 2000$ and $\sigma^2=4$,
Example \ref{exm4.2} with $n=500$ and $\sigma^2=4$,
and Example \ref{exm4.3} with $n=500$ and $\sigma^2=e^{0.7}$.
\cite{Morikawa2016}'s algorithm either does not converge or
produces too big (greater than 5) or too small (less than 0)  mean estimates.
Throughout the simulation study,
the performance of $\tilde \mu_t$ are evaluated based only
on the estimates between 0 and 5.

\begin{table}[!ht]
\renewcommand{\arraystretch}{0.66}
\caption{
Relative bias (RB; $\times   100$) and mean square error (MSE; $\times   100$) of five estimates of $\mu$.
\label{RB.MSE}
}
\centering
{
\tabcolsep 4pt
\vspace{0.1in}
\begin{tabular}{cc|ccccc|ccccc}   \hline
$n$&&$\hat\mu$& $\tilde \mu_{t}$ & $\tilde \mu_{np}$ & $\bar y_r$ & $\bar y$ &
   $\hat\mu$&$\tilde \mu_{t}$&$\tilde \mu_{np}$&$\bar y_r$&$\bar y$\\  \hline
 &&\multicolumn{5}{c|}{Example \ref{exm4.1}: $\sigma^2=1$}&\multicolumn{5}{c}{Example \ref{exm4.1}: $\sigma^2=4$}\\
500   &RB    &-0.12&-0.39&-1.31&-32.61&-0.19  &  0.35&-1.18&-8.45&-51.71&-0.19\\
500   &MSE   &0.93 & 0.98& 1.04& 33.10&0.81   &  4.00& 7.17& 7.18&146.23&1.78\\
2000  &RB    &0.10 & 0.04&-0.33&-32.54&0.03   &  0.18&-0.11&-4.21&-51.57&0.01\\
2000  &MSE   &0.22 & 0.24& 0.24& 32.49&0.19   &  0.98& 1.24& 1.91&144.24&0.44\\\hline
 &&\multicolumn{5}{c|}{Example \ref{exm4.2}: $\sigma^2=1$}&\multicolumn{5}{c}{Example \ref{exm4.2}: $\sigma^2=4$}\\
500   &RB    &-0.15&-0.28&-4.49&-35.83&0.01   &0.18&-0.68&-18.62&-56.27&0.05\\
500   &MSE   & 1.09& 1.12& 1.62& 35.49&0.93   &3.97& 5.68& 19.45&152.19&1.90\\
2000  &RB    & 0.14& 0.13&-2.85&-35.69&0.11   &0.15& 0.09&-15.01&-56.05&0.09\\
2000  &MSE   & 0.26& 0.27& 0.48& 34.43&0.23   &0.97& 1.40& 11.43&149.48&0.47\\  \hline
 &&\multicolumn{5}{c|}{Example \ref{exm4.3}: $\sigma^2=1$}&\multicolumn{5}{c}{Example \ref{exm4.3}: $\sigma^2=e^{0.7}$}\\
500   &RB    & 0.01&-0.24&-2.29&-18.70&0.06    &0.02&-0.71&-3.08&-23.00&0.06\\
500   &MSE   & 1.01& 1.82& 1.48& 34.79&0.90    &1.59&3.99&2.52&58.19&1.21\\
2000  &RB    & 0.05&-0.02&-1.20&-18.70&-0.04   &0.05&-0.11&-1.67&-23.06&0.00\\
2000  &MSE   & 0.25& 0.38& 0.39& 34.35&0.23    &0.41&0.80&0.69&57.76&0.29\\  \hline
\end{tabular}
}
\end{table}

We first look at  the results for Example \ref{exm4.1}.
When $\sigma^2=1$,  the relative biases of the proposed estimator
and \cite{Morikawa2016}'s two adaptive estimators are  all small.
The proposed estimator has slightly smaller mean square errors
than \cite{Morikawa2016}'s two adaptive estimators,
whose mean square errors are quite close to each other.
When $\sigma^2$ is increased to 4,
the relative biases of $\tilde \mu_{np}$ become much bigger.
The proposed estimator has much smaller mean square errors than
\cite{Morikawa2016}'s two adaptive estimators. 	
The comparison between $\hat\mu$ and $\tilde \mu_{t}$ in Example \ref{exm4.2} is similar
to that for Example \ref{exm4.1}.
For Example \ref{exm4.2}, compared with $\tilde \mu_{t}$,
$\tilde \mu_{np}$ has much bigger relative biases and mean square errors
especially for larger $\sigma^2$.
Next, we  examine the results for Example \ref{exm4.3},
in which there is no instrumental variable.
The proposed estimator has small relative biases in all situations.
Its mean square errors are significantly smaller
than \cite{Morikawa2016}'s two adaptive estimators.
Finally, as expected,  $\bar y_r$ has large relative biases
and the largest mean square errors in all examples,
while the ideal estimator $\bar y$ has small relative biases
and the smallest mean square errors in all situations.
When $\sigma^2$ is small, the proposed estimator has almost the same performance
as the ideal estimator $\bar y$, indicating that it is nearly optimal and
can be hardly  improved.

\subsection{Interval estimation}
This section is devoted to  comparing the coverage of Wald confidence intervals based on
$\hat\mu$, $\tilde\mu_t$, and $\bar y_r$.
The nonparametric bootstrap method with 200 bootstrap samples is used to
estimate the asymptotic variance for each of the three mean estimators.
  Although the variance estimator in \eqref{sigma-hat}
can be used in the  Wald-type confidence intervals based on $\hat\mu$,
its complicated form makes it more difficult to calculate
than bootstrap variance estimate.
The bootstrap method is quite computationally intensive for $\tilde\mu_{np}$.
For example, in Example \ref{exm4.1}, it takes around 9 minutes and 2 hours respectively to
calculate the bootstrap variances for $\tilde\mu_{np}$
for a single replication when $n=500$ and $n=2000$.
Hence we do not include it for comparison.
Again the number of repetitions is 2000 in all cases.
The simulation results are summarized in Table \ref{true.ci}.

In most cases, both  Wald confidence intervals based on
$\hat\mu$ and $\tilde\mu_t$ have very close
and accurate coverage probabilities.
The exceptions are Example 1 and Example 3 with the smaller sample size $n=500$.
For Example 1, both intervals have slight under coverage,
while for Example 3, the Wald confidence interval  based on
  $\tilde\mu_t$  has much over-coverage, in particular when
$\sigma^2$ is large.
When the sample size $n$ is increased to 2000,
both intervals have perfect coverage accuracy.
It is worth noting that  for the Wald confidence interval  based on
$\tilde\mu_t$,
the results with   $\tilde\mu_t$ outside $[0, 5]$
or not convergent were not taken into consideration.
In all cases, the Wald confidence interval  based on
$\bar y_r$ has unacceptable coverage accuracy,
which is most probably caused by the severe bias of $\bar y_r$.
Overall the Wald confidence interval  based on the proposed estimator
$\hat \mu$  is the most accurate and desirable among the three interval
estimators under comparison.

\begin{table}[!ht]
\renewcommand{\arraystretch}{0.66}
\caption{
Simulated coverage probabilities (\%) of bootstrap Wald-type
confidence intervals based on  $\hat\mu$, $\tilde\mu_t$,
and $\bar y_r$ in Examples \ref{exm4.1}--\ref{exm4.3}.
 \label{true.ci}
}
\centering
{
 \tabcolsep 4pt
\vspace{0.1in}
\begin{tabular}{clllll | clllll} \hline
Example &$\sigma^2$&$n$&$\hat\mu$&$\tilde\mu_{t}$ &$\bar y_r$
&Example &$\sigma^2$&$n$&$\hat\mu$&$\tilde\mu_{t}$ &$\bar y_r$\\\hline
1&1&500  &93.6   &94.4 & 0 &  1&4           &500  &95.1 &95.7 &1.0             \\
1&1&2000 &95.3   &95.1 & 0 &  1&4           &2000 &94.7 &94.2 &0               \\
2&1&500  &94.5   &94.7 &0.1&  2&4           &500  &95.2 &95.3&0              \\
2&1&2000 &95.1   &95.2 & 0 &  2&4           &2000 &95.4 &95.2 &0        \\

3&1&500  &94.9   &96.0 & 0 &  3&  $e^{0.7}$ &500  & 95.7&97.5&0    \\
3&1&2000 &95.0   &94.7 & 0 &  3&  $e^{0.7}$ &2000 & 94.8&95.5&0         \\ \hline
\end{tabular}
}
\end{table}

\section{An application}


We   apply the proposed method to analyze
the Korean Labor and Income Panel Study (KLIPS) data set
\citep{Kim2011, Wang2014, Shao2016, Morikawa2016},
which includes $n$ =2506 regular wage earners.
We take the monthly income in 2006 as the response $y$,
and use as covariates the monthly
income in 2005 ($x_1$), age ($x_2$), education level ($x_3$),   and gender ($x_4$).
The variable $y$ has approximately 35\% missing values,
while all the covariate values are observed.
The only outlier value of the education level, 99, is replaced by  6,
which is the nearest integer to the observed mean of the response $y$.
We treat all the variables  as continuous and suppose
  $\pr(D=0|\bx, y) = 1/\{1+\exp(\alpha^*+ \bx^\T \beta + y \gamma)\}$
  and that
$f(y|\bx, \xi)$ is  the density function of $N\big(\mu(\bx,\xi), \sigma^2(\bx,\xi)\big)$.
We consider the following four scenarios of $\bx$,  $\mu(\bx,\xi)$ and $\sigma^2(\bx,\xi)$:

\begin{description}
\item
Model (1a):
We take  $x=x_1$   and assume
  $\mu(x, \xi) =  \xi_0+\xi_1 x+\xi_2 x^2$ and $\sigma^2(x,\xi)=\exp(\xi_3+\xi_4 x)$.

\item
Model (1b):
We take $\bx= (x_1, x_2)^\T$   and assume
  $\mu(\bx,\xi) =  \xi_0 + \xi_1 x_1 +\xi_2 x_2
   + \xi_3 x_1^2 +
  \xi_4 x_1 x_2
  + \xi_5 x_2^2     $
and $\sigma^2(\bx,\xi)=\exp(\xi_{6}+\xi_{7} x_6 +\xi_{8} x_7 )$.

\item
Model (1c):
We take $\bx= (x_1, x_2, x_3)^\T$  and assume
  $\mu(\bx,\xi) =  \xi_0 + \xi_1 x_1 +\xi_2 x_2 + \xi_3 x_3
   + \xi_4 x_1^2 +
  \xi_5 x_1 x_2 + \xi_6 x_1 x_3
  + \xi_7 x_2^2  + \xi_8 x_2x_3
  + \xi_{9} x_3^2  $
and $\sigma^2(\bx,\xi)=\exp(\xi_{10}+\xi_{11} x_1 +\xi_{12} x_2 + \xi_{13} x_3)$.

  \item
Model (1d):
We take $\bx= (x_1, x_2, x_3, x_4)^\T$    and assume
  $\mu(\bx,\xi) =  \xi_0 + \xi_1 x_1 +\xi_2 x_2 + \xi_3 x_3
  + \xi_4 x_4 + \xi_5 x_1^2 +
  \xi_6 x_1 x_2 + \xi_7 x_1 x_3
  + \xi_8 x_2^2  + \xi_9 x_2x_3
  + \xi_{10} x_3^2  $
and $\sigma^2(\bx,\xi)=\exp(\xi_{11}+\xi_{12} x_1 +\xi_{13} x_2 + \xi_{14} x_3
  + \xi_{15} x_4)$.
\end{description}

Table \ref{tab-real1} presents $\hat \mu$, $\tilde \mu_t$, and $\bar y_r$,
and the corresponding Wald-type interval estimates at the 95\% confidence level.
Their standard errors are estimated by  the nonparametric bootstrap method
with 200 bootstrap samples.
The proposed mean estimates are all around 188 and the corresponding interval
estimates are around [177, 197].
In comparison,  \cite{Morikawa2016}'s point and interval estimates are
around 190 and [183, 197], except for model (1d), where
the interval estimate [174, 205] is much wider.
  Meanwhile we have observed that \cite{Morikawa2016}'s method suffers from
numerical problems.
Under the four models,  their algorithm did not converge
in 6.5\%,  28.5\%, 32.5\%, and 30\%, respectively, of the 200 bootstrap repetitions.
In the data-set under study,   the observations for the response $y$
are actually available,  but made missing by \cite{Kim2011}
in their empirical study. The true mean of $y$ is 185.04, which
is trustable to be close to the population mean.
The above observations indicate that the proposed method  has less bias in point estimation
and  more stable performance in interval estimation  than  \cite{Morikawa2016}'s  method.
Again  the complete-case estimate $\bar y_r$ is unacceptably biased.

Based on the proposed full log-likelihood,
the Bayesian Information Criteria (BIC) of the four models are
5393.443, 5375.292, 5336.675, and 5271.354, respectively,
which suggests model (1d) is the best.
We would recommend the use of model (1d) and estimate the population mean by
187.76, which is the closest to the ideal sample mean 185.04  among all estimates.

%

\begin{table}[!ht]
\renewcommand{\arraystretch}{0.66}
\caption{
Point and interval estimates of the response mean   for the KLIPS data. 
\label{tab-real1} }
 \vspace{0.1in}
\tabcolsep 4pt
\centering
\begin{tabular}{cccc }   \hline
 \multicolumn{4}{ c  }{KLIPS data }   \\ 
 Model &  Method &  Estimate     & Interval Estimate    \\      \hline
    &  $\bar y_r$       & 205.71 &  [200.12,  211.29]   \\
 1a &  $\hat \mu$       & 189.32 &  [180.68,  197.95]   \\
    &  $\tilde \mu_t$   & 190.25 &  [183.79,  196.71]   \\
 1b &  $\hat \mu$       & 189.15 &  [180.78,  197.52]   \\
    &  $\tilde \mu_t$   & 190.16 &  [181.90,  198.43]   \\
 1c &  $\hat \mu$       & 187.88 &  [178.90,  196.86]   \\
    &  $\tilde \mu_t$   & 191.38 &  [184.51,  198.25]   \\
 1d &  $\hat \mu$       & 187.76 &  [177.28,  198.25]   \\
    &  $\tilde \mu_t$   & 189.61 &  [171.10,  208.12]   \\     \hline


\end{tabular}
\end{table}

\section{Discussion}
In the development of our estimation procedure,  we assume that
the propensity score follows  a  logistic regression model only for convenience,
since the logistic regression model has a close relationship with \cite{Anderson1979}'s DRM.
Our estimation procedure for  the propensity score works for any parametric model
provided it is identifiable.

\setcounter{equation}{0}
\setcounter{section}{0}
\renewcommand{\theequation} {A.\arabic{equation}}
\section*{Appendix 1: Regularity conditions for $f(y|\bx,\xi)$}

The following regularity conditions on $f(y|\bx,\xi)$ mimic those
for the consistency and asymptotic normality of the MLE
under a regular parametric model on pp.~144--145 of  Serfling (1980).

\ben

\item[(A1)]
In a neighbourhood of $ \xi_0 $,
$ \log \{ f(y|\bx, \xi) \} $ is  three-times differentiable with respect to $\xi$
for any $(y, \bx)$.

\item[(A2)]

 For $(\gamma, \xi)$ in a neighbourhood of $(\gamma_0, \xi_0)$
 and any $\bx$ on S,
 the inequality
$\int e^{y\gamma}  f(y|\bx, \xi) dy <\infty$ holds.

 \item[(A3)]
The matrix $V$ {defined in (\ref{var.mat})}  is well defined and nonsingular.

\item[(A4)]
{There exists a function $M(\bx)$ not depending on $(\gamma, \xi)$ such that
$
\e \{ M( \bX) \} <\infty
$
and
\bas
\left\| \int  e^{y\gamma} \nabla_{\xi}  f(y|\bx, \xi) dy \right \|
+\left \| \int e^{y\gamma} \nabla_{\xi,\xi }  f(y|\bx, \xi) dy  \right\| +
\left\| \int e^{y\gamma} \nabla_{\xi,\xi,\xi} f(y|\bx, \xi) dy \right \|  < M( \bx)
\eas
uniformly for $(\gamma, \xi)$ in  a neighbourhood of $(\gamma_0, \xi_0)$
and a neighbourhood of $( 0, \xi_0)$.
Here
$\nabla_{\xi}  f(y|\bx, \xi)$, $\nabla_{\xi,\xi}  f(y|\bx, \xi)$,
and $\nabla_{\xi,\xi,\xi}  f(y|\bx, \xi)$ are the first, second,
and third derivatives of $f(y|\bx, \xi)$ with respect to $\xi$.

}

\een
Conditions A1 and A2 guarantee that  the first partial derivatives of
$\ell_2(\alpha, \beta, \xi, \gamma)$ are well defined
for any data.
Conditions  A3 and A4 ensure that
$\ell_2(\alpha, \beta, \xi, \gamma)$  can be approximated by
its second-order Taylor expansion at
$ (\alpha_0, \beta_0, \xi_0, \gamma_0)$
when $(\alpha, \beta, \xi, \gamma)$ lies in a neighbourhood of
$ (\alpha_0, \beta_0, \xi_0, \gamma_0)$.

\section*{Appendix 2: Proof of Corollary  1}

We first consider Part (a).
Since $z$ is an instrument variable, (\ref{drm2}) becomes
\bas
\pr(\bx |D=0)
&=& \exp\{ \alpha + u^\T\beta+c(\bx,  \gamma,\xi) \} \pr(\bx |D=1).
\eas
The  identification of $(\alpha, \beta, \gamma)$  is equivalent to
the identification of $(\alpha, \beta, \gamma)$
in $ \alpha + u^\T\beta+c(\bx,  \gamma,\xi) $.

Recall that $f(y|\bx, \xi)=\pr(y|\bx, D=1)$. Then
\bas
f (y|\bx, \xi)
&=& \pr(y|\bx) \pr(D=1|\bx, y)\Big/\int \pr(y|\bx) \pr(D=1|\bx, y) dy  \\
&=& \pr(y|z, u) \pr(D=1|u, y)\Big/\int \pr(y|z, u) \pr(D=1|u, y) dy.
\eas
Since $z$ is an instrument variable,
it follows that $f (y|\bx, \xi)$ must depend on $z$,
and so must $c(\bx, \gamma,\xi) $.
Suppose
$(\alpha_1, \beta_1, \gamma_1)$ and $(\alpha_2, \beta_2, \gamma_2)$
satisfy
\bas
\alpha_1 + u^\T\beta_1+c(\bx,  \gamma_1,\xi)
=\alpha_2 + u^\T\beta_2+c(\bx,  \gamma_2,\xi)
\eas
for all $\bx$,
which implies
\bas
 c(\bx,  \gamma_1,\xi_0)
-c(\bx,  \gamma_2,\xi_0)
=
 (\alpha_2 - \alpha_1) + u^\T(\beta_2-\beta_1).
\eas
Since the left-hand side depends on $z$, while the right-hand side does not,
we must have $\gamma_1=\gamma_2$,
which further implies that $\alpha_1=\alpha_2$
and $\beta_1=\beta_2$.
This indicates that  the parameters $(\alpha, \beta, \gamma)$ are identifiable,
which completes the proof of Part (a).

We next consider Part (b). Suppose  $(\alpha_1, \beta_1, \gamma_1)$
and  $(\alpha_2, \beta_2, \gamma_2)$ satisfy
\bas
\alpha_1 + \beta_1^\T \bx + c(\bx,  \gamma_1, \xi_0)
=
\alpha_2 + \beta_2^\T \bx + c(\bx,  \gamma_2, \xi_0)
\eas
for all $\bx \in S$.
According to the expression for $c(\bx,  \gamma, \xi_0)$,
this implies that
\bas
&&  \sum_{i=1}^{k}  a_{i}(\gamma_1) g_i(\bx) + \{\alpha_1+ a_{k+1}(\gamma_1)\}
   + \bx^\T \{ \beta_1 +  a_{k+2}(\gamma_1) \}  \\
&=&
  \sum_{i=1}^{k}  a_{i}(\gamma_2) g_i(\bx) + \{\alpha_2+ a_{k+1}(\gamma_2)\}
   + \bx^\T \{ \beta_2+ a_{k+2}(\gamma_2)\}.
\eas
Since  $1, \bx, g_1(\bx), \ldots, g_{k}(\bx)$  are linearly independent,
it follows that
\bas
\big( a_{1}(\gamma_1), \ldots, a_{k}(\gamma_1)  \big)
&=& \big( a_{1}(\gamma_2), \ldots, a_{k}(\gamma_2)  \big),
\\
 \alpha_1+ a_{k+1}(\gamma_1)  &=&    \alpha_2+ a_{k+1}(\gamma_2),  \\
\beta_1 +   a_{k+2}(\gamma_1)
&=& \beta_2+  a_{k+2}(\gamma_2)
\eas
hold  simultaneously.   Because
 $\big(a_1(\gamma_1), \ldots, a_{k}(\gamma_1)\big) \neq \big(a_1(\gamma_2), \ldots, a_{k}(\gamma_2)\big)$ for any $\gamma_1\neq \gamma_2$,
 the first equation implies $\gamma_1=\gamma_2$.
Then the last two equations lead to $\alpha_1=\alpha_2$
and $\beta_1=\beta_2$.
This completes the proof of Part (b) and that of Corollary  1.

\section*{Appendix 3: Preparation for proving Theorems 1--3}

\subsection*{Re-expression}

It can be verified that
$$
\ell_2(\theta) =  h(\theta, \lambda_{\theta}),
$$
where
\ba
h(\theta, \lambda)
&=&   \sum_{i=1}^{n_1}\log \{  f(y_i|\bx_i, \xi)\} + \sum_{i=n_1+1}^n t(\bx_i,\theta) \nonumber \\
&&
- \sum_{i=1}^{n} \log\{1+ \lambda [\exp\{t(\bx_i,\theta) \}-1]\}
\label{lik.part2}
\ea
and
$\lambda_{\theta}$ is the solution to $ \nabla_{\lambda}  h  = 0$.

Let $\hat \lambda$  be the solution to (\ref{mle.lambda})  with $\hat\theta$ in place of $\theta$.
We first discuss some properties of $\hat\lambda$.
It can be verified that $(\hat\theta,\hat\lambda)$
satisfy
$$
\nabla_{\alpha}  h(\hat\theta,\hat\lambda) =0,~~
\nabla_{\lambda}  h(\hat\theta,\hat\lambda) =0.
$$
Note that {
\bas
\nabla_{\lambda}  h(\theta, \lambda)
&=&
-\sum_{i=1}^n  \frac{\exp\{ t(\bx_i,\theta)\}-1}{1+ \lambda[\exp\{t(\bx_i,\theta)\}-1]} = 0
\quad \text{ and}  \\
\nabla_{\alpha}  h(\theta, \lambda)
&=&n_2-\lambda \sum_{i=1}^n  \frac{\exp\{ t(\bx_i,\theta)\}}{1+ \lambda[\exp\{t(\bx_i,\theta)\}-1]} = 0
\eas}
together imply that
\begin{equation}
\label{lambda.equation.m1}
\hat\lambda=n_2/n,
\end{equation}
which converges in probability to $\lambda_0=1-\eta_0$.

For convenience of presentation, let
$
\omega= (\theta^\T, \lambda)^\T.
$
It can be verified that $\hat \omega= (\hat\theta^\T, \hat\lambda)^\T$ is the solution to
$\partial h(\theta,\lambda)/\partial \omega= 0$.
To investigate the asymptotic properties of  $(\hat\theta,\hat\lambda)$,
we need their approximations, which can be obtained via the second-order Taylor expansion
 of $h(\theta,\lambda)$ around $\omega=\omega_0\equiv (\theta_0^\T,\lambda_0)^\T$.
In the next subsection, we derive the forms of
${\partial h(\theta_0,\lambda_0)}/{\partial \omega}$ and
${\partial^2 h(\theta_0,\lambda_0)}/{\left(\partial \omega\partial \omega^{\T}\right)}$
 and study their properties.

\subsection*{First and second derivatives of $h(\theta,\lambda)$ at $\omega=\omega_0$ }

For convenience of presentation, we write $\pi_i=\pi(\bx_i)$.
Denote
\begin{equation}
\label{un.def}
u_n
=\left(u_{n1},u_{n2}^\T \right)^\T,
\end{equation}
where
\bas
u_{n1}
&=&\nabla_{\theta}  h(\theta_0,\lambda_0)
=
\sum_{i=1}^{n} \left[
(1-d_i - \pi_i ) \nabla_{\theta} t(\bx_i,\theta_0)
+ d_i I_e \nabla_{\xi}     \log \{f(y_i|\bx_i, \xi_0)\}\right],
\\
u_{n2}
&=&\nabla_{\lambda} h(\theta_0,\lambda_0)
=  \frac{1}{\lambda_0(1-\lambda_0)}\sum_{i=1}^{n}\left(\lambda_0-\pi_i\right).
\eas

After some calculation,
it can be verified that  the second derivatives of $h(\theta,\lambda)$ at $(\theta_0,\lambda_0)$
are
\bas
\nabla_{\theta\theta^\T} h(\theta_0,\lambda_0)
&=& V_n=
\sum_{i=1}^{n} d_i  I_e \nabla_{\xi\xi}
    \log \{f(y_i|\bx_i, \xi_0)\} I_e^\T \\
    && \hspace{1cm} + \sum_{i=1}^n(1-d_i-\pi_i)\nabla_{\theta\theta}  t(\bx_i,\theta_0)\\
    &&\hspace{1cm}- \sum_{i=1}^{n} \pi_i(1-\pi_i) \{ \nabla_{\theta}t(\bx_i,\theta_0)\}^{\otimes 2}, \\
\nabla_{\theta\lambda} h(\theta_0,\lambda_0)
&=& \frac{1}{\lambda_0(1-\lambda_0)}V_n\bfe_1,
\\
\nabla_{\lambda\lambda} h(\theta_0,\lambda_0)
&=& \frac{1}{\lambda^2_0(1-\lambda_0)^2}\sum_{i=1}^n(\lambda_0-\pi_i)^2.
\eas

\subsection*{Some useful technical lemmas}
When deriving the asymptotic distribution of $\hat\theta$, we need to use
$\e \{ \nabla_{\theta\theta^\T} h(\theta_0,\lambda_0) \}$, $\e \{ \nabla_{\theta\lambda} h(\theta_0,\lambda_0) \}$,
$\e \{ \nabla_{\lambda\lambda} h(\theta_0,\lambda_0) \}$,
and
the expectation and variance of $u_n$ defined in (\ref{un.def}).
We need the following lemma to simplify our calculation.

\begin{lemma}
\label{u.lemma1}
The following equations hold:
\ba
\e[  d_i \nabla_{\xi}     \log \{f(y_i|\bx_i, \xi_0)\} ] &=& 0, \label{eq1} \\
\e[ d_i   \nabla_{\xi\xi}
    \log \{f(y_i|\bx_i, \xi_0)\}  ]
&=&
- \e\{ d_i [ \nabla_{\xi }
    \log \{f(y_i|\bx_i, \xi_0)\} ]^{\otimes 2} \},  \label{eq2} \\
-\frac{1}{n}
\e \{ \nabla_{\theta\theta^\T} h(\theta_0,\lambda_0) \} &=&  V, \label{eq3} \\
-\frac{1}{n}
\e \{ \nabla_{\theta\lambda} h(\theta_0,\lambda_0) \} &=&   \frac{1}{\lambda_0(1-\lambda_0)}V \bfe_1, \label{eq4} \\
-\frac{1}{n}
\e \{ \nabla_{\lambda\lambda} h(\theta_0,\lambda_0) \} &=&
\frac{\bfe_1^\T V\bfe_1 - \lambda_0(1-\lambda_0)}{\lambda^2_0(1-\lambda_0)^2}.  \label{eq5}
\ea
\end{lemma}
\proof
By the fact $f(y|\bx, \xi) = \pr(Y=y|\bX=\bx, D=1)$, it can be verified that
\bas
\e[   \nabla_{\xi}
    \log \{f(y_i|\bx_i, \xi_0)\} | \bx_i, d_i=1  ]
&=&
0, \\
\e[   \nabla_{\xi\xi}
    \log \{f(y_i|\bx_i, \xi_0)\} | \bx_i, d_i=1  ]
&=&
-
\e\{  [ \nabla_{\xi }
    \log \{f(y_i|\bx_i, \xi_0)\}]^{\otimes 2} | \bx_i, d_i=1 \},
\eas
which imply respectively Equations\eqref{eq1} and \eqref{eq2} by conditioning on $(\bx_i, d_i=1)$.

Equations \eqref{eq3} and \eqref{eq4} follows immediately  from \eqref{eq2}.
To prove \eqref{eq5}, by noticing
\[
\lambda_0 = 1-\eta_0 = \pr(D=0)  \quad \mbox{and}\quad
\pi(\bx) = \pr(D=0|\bx),
\]
we have
  $\e  \{ \pi(\bX)\} =  \lambda_0$
and
\ba
\label{eq5-im}
\frac{1}{n}
\e \{ \nabla_{\lambda\lambda} h(\theta_0,\lambda_0) \} &=&
 \e\{\lambda_0 - \pi(\bx_i)\}^2  = \e [ \{ \pi(\bX)\}^2 ]  - \lambda_0^2.
\ea
Since $\bfe_1^\T V\bfe_1 =   \e[ \pi(\bX) \{1-\pi(\bX)\}]  =   -  \e [ \{ \pi(\bX)\}^2 ]  + \lambda_0 $,
Equation \eqref{eq5} follows by comparing
this equation with  \eqref{eq5-im}.
This finishes the proof.
\qed

The final lemma presents the expectation and variance of $u_n$.
\begin{lemma}
\label{u.lemma2}
With $u_n$ defined in (\ref{un.def}), we have
$
\e(u_n)=0
$
and
$$
\frac{1}{n}\var(u_n)=U=
\left(
\begin{array}{cc}
V&0\\
0&\frac{ -\bfe_1^\T V\bfe_1 + \lambda_0(1-\lambda_0)}{\lambda_0^2(1-\lambda_0)^2}
\end{array}
\right).
$$
\end{lemma}
\proof

The result $\e (u_{n2})=0$ follows from
$\pi(\bx) = \pr(D=0|\bX=\bx)$  and  Equation \eqref{eq1}.

For $\var(u_n)$, we first calculate $\var(u_{n2})$.
It can be seen that
\bas
\frac{1}{n} \var(u_{n2})
&=&
\frac{1}{\lambda_0^2 (1-\lambda_0)^2} \e[\{\lambda_0-\pi(\bx_i) \}^2]
=
\frac{1}{\lambda_0^2 (1-\lambda_0)^2} [ \e\{ \pi(\bx_i) \}^2 - \lambda_0^2].
\eas
We have shown that $\bfe_1^\T V\bfe_1   = -  \e [ \{ \pi(\bX)\}^2 ]  + \lambda_0$
in the proof of Lemma 1.
Therefore
\bas
\frac{1}{n} \var(u_{n2}) = \frac{  -\bfe_1^\T V\bfe_1 + \lambda_0(1-\lambda_0)}{\lambda_0^2(1-\lambda_0)^2}.
\eas

It remains to calculate  $\var(u_{n1})$.
Re-write
\[
u_{n1} =\sum_{i=1}^{n} ( u_{n11, i} + u_{n12, i}),
\]
where
\bas
u_{n11, i}
&=&
(1-d_i - \pi_i ) \nabla_{\theta} t(\bx_i,\theta_0) \\
u_{n12, i}
&=&  d_i I_e \nabla_{\xi}     \log \{f(y_i|\bx_i, \xi_0)\}.
\eas
Since both $u_{n11, i}$ and $ u_{n12, i}$  have mean zero,
it follows from equality \eqref{eq1} that
\bas
\frac{1}{n}\var(u_{n1})
&=&
\e \{ ( u_{n11, i} + u_{n12, i} )^{\otimes 2}\}
= \e ( u_{n11, i}^{\otimes 2})  + \e  ( u_{n12, i}^{\otimes 2}).
\eas
Because $\e(d_i=0|\bx_i) = \pr(D=0|\bX=\bx_i) = \pi(\bx_i)$,
by conditioning on $\bx_i$, we have
\bas
\e ( u_{n11, i}^{\otimes 2})
&=&
\e[ \{ (1-d_i - \pi_i ) \nabla_{\theta} t(\bx_i,\theta_0) \}^{\otimes 2} ] \\
&=&
\e  [  (\pi(\bx_i)\{1-\pi(\bx_i)\}  \{\nabla_{\theta} t(\bx_i,\theta_0) \}^{\otimes 2}].
\eas
Clearly
$
\e  ( u_{n12, i}^2) = \e[  d_i I_e \nabla_{\xi}     \log \{f(y_i|\bx_i, \xi_0)\} ]^2.
$
This proves $\frac{1}{n}\var(u_{n1}) = V$ by
 comparing the expression of $V$ with $\frac{1}{n}\var(u_{n1})$.
\qed

\section*{Appendix 4: Proof of Theorem  1}
We start with Part (a).
Using a similar argument to that used in the proofs of Lemma 1
and Theorem 1 of \citet{Qin1994}, we have
$
\hat\theta=\theta_0+O_p(n^{-1/2})
$
and $\hat\lambda-\lambda_0=O_p(n^{-1/2})$.
Next we investigate the asymptotic approximation of $\hat\theta$.

The maximum likelihood estimator $\hat\theta$ of $\theta$
and the associated Lagrange multiplier $\hat \lambda$
must satisfy
$$
\left(
\begin{array}{c}
\nabla_{\theta} h(\hat\theta,\hat \lambda)  \\
\nabla_{\lambda} h(\hat\theta,\hat \lambda)
\end{array}
\right)
=0.
$$
Applying a first-order expansion  to the left-hand side of the above equation
gives
\begin{equation}
\label{score}
0=
\left(
\begin{array}{c}
\nabla_{\theta} h(\theta_0, \lambda_0)  \\
\nabla_{\lambda} h(\theta_0, \lambda_0)
\end{array}
\right)
+
\left(
\begin{array}{cc}
\nabla_{\theta\theta^\T} h(\theta_0, \lambda_0)  & \nabla_{\theta\lambda} h(\theta_0, \lambda_0)\\
\nabla_{\lambda \theta^\T} h(\theta_0, \lambda_0) & \nabla_{\lambda \lambda} h(\theta_0, \lambda_0)
\end{array}
\right)
\left(
\begin{array}{c}
\hat \theta-\theta_0  \\
\hat \lambda - \lambda_0
\end{array}
\right) + o_p(n^{1/2}).
\end{equation}
By Lemma 1,
\begin{equation}
\label{el2.omega}
\left(
\begin{array}{cc}
\nabla_{\theta\theta^\T} h(\theta_0, \lambda_0)  & \nabla_{\theta\lambda} h(\theta_0, \lambda_0)\\
\nabla_{\lambda \theta^\T} h(\theta_0, \lambda_0) & \nabla_{\lambda \lambda} h(\theta_0, \lambda_0)
\end{array}
\right)
=-n
W
+o_p(n),
\end{equation}
where
\bas
W =
\left(
\begin{array}{cc}
V  &  \frac{1}{\lambda_0(1-\lambda_0)}V \bfe_1 \\
 \frac{1}{\lambda_0(1-\lambda_0)} \bfe_1^\T  V &
  \frac{\bfe_1^\T V\bfe_1 -  \lambda_0(1-\lambda_0)}{\lambda^2_0(1-\lambda_0)^2}
\end{array}
\right).
\eas
Recall that $u_n=(\nabla_{\theta} h(\theta_0,\lambda_0), \nabla_{\lambda}h(\theta_0,\lambda_0) )$.
Combining (\ref{score}) and (\ref{el2.omega}), we get
\begin{equation}
\label{mle.approx1}
\left(
\begin{array}{c}
\hat \theta-\theta_0  \\
\hat \lambda - \lambda_0
\end{array}
\right)
=\frac{1}{n}W^{-1}u_n+o_p(n^{-1/2}).
\end{equation}
{Note that
$|W| = - |V|\cdot |\lambda_0(1-\lambda_0)| = - |V|\cdot |\eta_0(1-\eta_0)|$
and we have assumed that $\eta_0\in (0, 1)$ and $V$ is nonsingular,
the matrix $W$ is nonsingular and its inverse $W^{-1}$ is well defined. }
Since
\begin{eqnarray}
\label{W-inv}
W^{-1}
=\left(
\begin{array}{ccc}
V^{-1}-\frac{1}{\lambda_0(1-\lambda_0)}\bfe_1\bfe_1^\T&\bfe_1\\
\bfe_1^\T&-\lambda_0(1-\lambda_0)\\
\end{array}
\right),
\end{eqnarray}
we have
\begin{equation}
\label{mle.theta}
\hat \theta-\theta_0=
n^{-1}\left(
\begin{array}{ccc}
V^{-1}-\frac{1}{\lambda_0(1-\lambda_0)}\bfe_1\bfe_1^\T\quad &\bfe_1\\
\end{array}
\right)u_n+o_p(n^{-1/2}).
\end{equation}
With Lemma \ref{u.lemma2},
we can verify that
$$
\var\left\{n^{-1/2}\left(
\begin{array}{ccc}
V^{-1}-\frac{1}{\lambda_0(1-\lambda_0)}\bfe_1\bfe_1^\T\quad &\bfe_1\\
\end{array}
\right)u_n\right\}
=V^{-1}-\frac{1}{\lambda_0(1-\lambda_0)}\bfe_1\bfe_1^\T.
$$
Note that $u_n$ is the sum of independent and identically distributed random vectors.
Hence,
$$
\sqrt{n}(\hat \theta-\theta_0)
\to N\Big(0, \quad V^{-1}-\frac{1}{\lambda_0(1-\lambda_0)}\bfe_1\bfe_1^\T\Big)
$$
in distribution. This completes the proof of Part (a).

Next, we consider Part (b). Recall that
$
R(\theta)
=
2\{  \ell_2(\hat \theta) -  \ell_2(\theta )  \}.
$
Then
$
R(\theta_0)
=2\{h(\hat\theta,\hat\lambda)-h(\theta_0,\lambda_{\theta_0})\},
$
where
$\lambda_{\theta_0}$ is the solution to $ \partial h(\theta_0,\lambda)/\partial \lambda = 0$.

Applying a second-order   Taylor  expansion to
$h(\hat\theta,\hat \lambda)$ and using (\ref{mle.approx1}),
we have
\begin{equation}
\label{lrt.part1}
h(\hat\theta,\hat\lambda)=\frac{n}{2}u_n^\T W^{-1} u_n+o_p(1).
\end{equation}
Following a similar argument to that for (\ref{lrt.part1}),
we get
\begin{equation}
\label{lrt.part2}
   h(\theta_0, \lambda_{\theta_0})
=-\frac{n}{2}u_{n2}^2\frac{\lambda_0^2(1-\lambda_0)^2}{\lambda_0(1-\lambda_0)- \bfe_1^\T V  \bfe_1}+o_p(1).
\end{equation}

Combining  (\ref{lrt.part1}) and (\ref{lrt.part2}) gives
\begin{eqnarray}\nonumber
R(\theta_0)=
nu_n^{\T}
\left(
\begin{array}{ccc}
V^{-1}-\frac{\bfe_1\bfe_1^\T}{\lambda_0(1-\lambda_0)}&\bfe_1\\
\bfe_1^\T&   \frac{\lambda_0(1-\lambda_0)\bfe_1^\T V  \bfe_1}{\lambda_0(1-\lambda_0)-\bfe_1^\T V  \bfe_1}\\
\end{array}
\right)u_n+o_p(1).
\end{eqnarray}
{Since  $W^{-1}$ are invertible,
the matrix $V^{-1} - \{\lambda_0(1-\lambda_0)\}^{-1}\bfe_1\bfe_1^\T$ is also invertible.
}
Let
$$
v_n=
u_{n1}+\left[V^{-1} - \{\lambda_0(1-\lambda_0)\}^{-1}\bfe_1\bfe_1^\T\right]^{-1}\bfe_1u_{n2}.
$$
After some algebra, $R(\theta_0)$ can be written as
$$
R(\theta_0)
=nv_n^\T\left[V^{-1}- \{\lambda_0(1-\lambda_0)\}^{-1}\bfe_1\bfe_1^\T\right] v_n+o_p(1).
$$
With Lemma \ref{u.lemma2},
we can further verify that  $\e(v_n)=0$
and
$$
\var\left(n^{-1/2}v_n\right)
=V+\frac{V\bfe_1\bfe_1^\T V}{\lambda_0(1-\lambda_0)- \bfe_1^\T V  \bfe_1 }
=
\left[ V^{-1}- \{\lambda_0(1-\lambda_0)\}^{-1}\bfe_1\bfe_1^\T\right]^{-1}.
$$
Hence, {
$
R(\theta_0)\to  \chi^2_{d_{\theta}}
$
in distribution. }
This completes the proof of Theorem  1 in the main paper.

\section*{Appendix 5: Proof of Theorem  2}

Recall that $\hat \eta = n_1/n = 1-\hat \lambda$ with $\eta_0 = \pr(D=1)$.
Then $\hat\mu$ in (\ref{mle1}) can be rewritten as
\bas
\hat \mu &=&
\sum_{i=1}^n   \hat p_i
\left[ \int   y \{ \hat \eta +(1-\hat \eta)\exp(\hat \alpha + \bx_i^\T\hat \beta  +\hat \gamma y) \} f(y|\bx_i, \hat \xi)dy\right] \\
   &=&
  \frac{1}{n} \sum_{i=1}^n \frac{  \int   y \{  \hat \eta  + (1-\hat \eta)\exp(\hat \alpha + \bx_i^\T\hat \beta  +\hat \gamma y) \} f(y|\bx_i, \hat \xi)dy  }{
 \hat \eta  + (1-\hat \eta) \exp\{ \hat \alpha +\bx_i^\T\hat \beta+c(\bx_i,\hat \gamma,\hat \xi)\} } \\
&=& n^{-1} \sum_{i=1}^n K(\bx_i;\hat \theta, \hat \eta),
\eas
where
\bas
K(\bx;\theta,\eta)&=&
 \frac{
\int  y \{  \eta + (1-\eta)\exp(  \alpha + \bx^\T \beta  + \gamma y) \} f(y|\bx,   \xi)dy  }{
\eta + (1-\eta) \exp\{  \alpha +\bx^\T  \beta+c(\bx, \gamma,  \xi)\} }.
\eas

Applying the first-order Taylor  expansion and the law of large numbers,
we have
$$
\hat\mu=\frac{1}{n}\sum_{i=1}^nK(\bx_i;\theta_0,\eta_0)+ A^\T(\hat\theta-\theta_0)
- B\{(1-\hat \eta)- (1-\eta_0) \} +o_p(n^{-1/2}),
$$
where $A=\e\left\{ \nabla_{\theta}K(\bX;\theta_0,\eta_0)\right\}$
and
$B=\e\left\{ \nabla_{\eta}K(\bX;\theta_0,\eta_0)\right\}$.
Hence, with Equation \eqref{mle.approx1} and $\hat \eta=1-\hat \lambda$, we have
$$
\hat\mu-\mu
=\frac{1}{n}\sum_{i=1}^n\{K(\bx_i;\theta_0,\eta_0)-\mu\}+n^{-1}(A^\T, -B) W^{-1}u_n+o_p(n^{-1/2}).
$$

We first argue that $E\{K(\bX;\theta_0,\eta_0)\}=\mu$.
  By  \eqref{drm2},  we have
\begin{equation}
\label{e.k1}
\pr(\bx)= \{ \eta_0 + (1-\eta_0) \exp\{  \alpha_0 +\bx^\T  \beta_0+c(\bx, \gamma_0,  \xi_0)\}  \pr(\bx|D=1).
\end{equation}
It then  follows that
\bas
&& \e\{ K(\bX;\theta_0,\eta_0) \} \\ &=&
\int_{\bx}  \frac{
\int_y  y \{  \eta_0 + (1-\eta_0)\exp(  \alpha_0 + \bx^\T \beta_0  + \gamma_0 y) \} f(y|\bx,   \xi_0)dy  }{
\eta_0 + (1-\eta_0)\exp\{  \alpha_0 +\bx^\T  \beta_0+c(\bx, \gamma_0,  \xi_0)\} } \pr(\bx)d\bx\\
 &=&\int_{\bx}  \int_y  y \{  \eta_0 + (1-\eta_0)\exp(  \alpha_0 + \bx^\T \beta_0  + \gamma_0 y) \} f(y|\bx,   \xi_0) \pr(\bx|D=1) dyd\bx\\
 &=&\int_{\bx}  \int_y  y \{  \eta_0 + (1-\eta_0)\exp(  \alpha_0 + \bx^\T \beta_0  + \gamma_0 y) \}\pr(y, \bx|D=1) dyd\bx\\
  &=&
\int_{\bx, y}
 y   \pr(y, \bx   )dy   d\bx= \mu.
 \eas

After some calculus, we found that
$
- B
=
    A^\T \bfe_1/\{ (1-\eta_0)\eta_0 \}.
$
With \eqref{W-inv}, we have
\bas
&& (A^\T, -B )W^{-1}u_n  \\
&=& A^\T \left(I, \frac{\bfe_1}{\lambda_0(1-\lambda_0)} \right)
\left(
\begin{array}{ccc}
V^{-1}-\frac{1}{\lambda_0(1-\lambda_0)}\bfe_1\bfe_1^\T&\bfe_1\\
\bfe_1^\T&-\lambda_0(1-\lambda_0)\\
\end{array}
\right)
\left(
\begin{array}{c}
u_{n1}\\
u_{n2}\\
\end{array}
\right)  \\
&=& A^\T V^{-1} u_{n1}.
\eas
Since $\e( u_{n1} |\bx_1, \ldots, \bx_n  ) = 0$, we arrive at
\bas
&&\cov\left(\sum_{i=1}^n K(\bx_i;\theta_0,\eta_0), (A^\T, -B) W^{-1}u_n\right)  \\
&=&
\cov\left(\sum_{i=1}^n K(\bx_i;\theta_0,\eta_0), A^\T V^{-1} u_{n1}\right)   \\
&=&0.
\eas

Finally,
By  Lemma 2 and the central limit theorem and Slutsky's theorem, we have
$$
\sqrt{n}(\hat\mu-\mu)
\to N\left(0,\sigma^2 \right),
$$
where $\sigma^2 = \var\{K(\bX;\theta_0,\eta_0) \}+  A^\T V^{-1} A$.
This proves Theorem  2 of the main paper.

\section*{Appendix 6: Proof of Theorem 3}

\subsection*{Preparations}
The observed data are  $(d_i=1, \bx_i, y_i)$ ($i=1, \ldots, n_1$)
and $(d_i=0, \bx_i)$ ($i=n_1+1, \ldots, n$).
We make two parametric assumptions:
\bas
\pr(D=1|\bX=\bx, Y=y) &=& \frac{1}{1+\exp\{ \alpha^* + \bx^\T \beta + \gamma y\}}, \\
\pr(Y=y|\bX=\bx, D=1) &=& f(y|\bx, \xi).
\eas
Recall that $\eta = \pr(D=1)$ and
$c(\bx, \gamma, \xi)
= \log\int e^{\gamma y} f(y|\bx, \xi) dy$.
Let $\vartheta=(\beta, \gamma, \xi)$
and
$
r(\bx, \vartheta)
=
 \bx^\T \beta + c(\bx, \gamma, \xi),
$
so that $\theta=(\alpha, \vartheta^\T)^\T$
and $t(\bx, \theta) = \alpha + r(\bx, \vartheta)$.
We have shown
\bas
\pr(y, \bx|D=0) &=&  \exp\{ \alpha  + \bx^\T \beta + \gamma y\} \pr(y, \bx|D=1),\\
\pr(  \bX=\bx| D=0) &=& \exp\{ \alpha+ r(\bx, \vartheta) \} \pr(  \bX=\bx| D=1),
\eas
where $\alpha=\alpha^* + \log\{\eta/(1-\eta)\}$.

In addition,
\bas
\e(1-D|\bX=\bx) = \pr(D=0|\bX=\bx) = \pi(\bx; \alpha, \vartheta, \eta),
\eas
where we have defined
\bas
\pi(\bx; \alpha, \vartheta, \eta)
=  \frac{ (1-\eta) \exp\{ t(\bx, \theta) \} }{\eta + (1-\eta) \exp\{ t(\bx, \theta) \}}
= \frac{ (1-\eta) \exp\{ \alpha+ r(\bx, \vartheta) \} }{\eta + (1-\eta) \exp\{ \alpha+ r(\bx, \vartheta) \}}
\eas
with  $\pi(\bx)$ abbreviation for $   \pi(\bx; \theta_0, \eta_0)$.

The observed data are iid from $(D, \bX, \tilde Y)$,
where $\tilde Y$ is empty when $D=0$,
and $\tilde Y=Y$ when $D=1$.
The joint distribution of $(D, \bX, \tilde Y)$ is
\bas
&&
\{ \pr(Y=y|\bX=\bx, D=1)\pr(\bX=\bx| D=1) \pr(D=1)\}^d  \\
&&\hspace{1cm}\times  \{ \pr(\bX=\bx| D=0) \pr(D=0)   \}^{1-d}  \\
&=&
\{ \pr(Y=y|\bX=\bx, D=1) \pr(D=1) \}^d  \\
&&\hspace{1cm}
\times  \{ \exp(t(\bx, \theta)) \pr(D=0)   \}^{1-d} \times \pr(\bx|D=1) \\
&=&
\{f(y|\bx, \xi) \eta  \}^d
\times \{ \exp(\alpha+ r(\bx, \vartheta)) (1-\eta)  \}^{1-d} \times \pr(\bx|D=1).
\eas
Here all except $\pr( \bX=\bx|D=1 )$ are completely parametric, and
we regard  $\pr( \bX=\bx|D=1 )$ as an infinite-dimensional parameter,
or simply
\bas
\pr( \bX=\bx|D=1 )\geq 0,  \quad
\int \pr( \bX=\bx|D=1 ) d\bx =1.
\eas
Therefore our imposed model is clearly semi-parametric.

Throughout this section,
we use $g(\bx, \zeta)$ to denote a parametric submodel  for $\pr( \bX=\bx |D=1)$ with
 $g(\bx, \zeta_0)$ being the true model.
The joint  density function of $(D, \bX, \tilde Y)$ is
 \bas
  h(d, \bx, y; \alpha, \vartheta, \eta, \zeta)
&=&
\{f(y|\bx, \xi) \eta  \}^d
\times \{ \exp(\alpha+ r(\bx, \vartheta)) (1-\eta)  \}^{1-d} \times g(\bx, \zeta).
\eas
It is worth noting that $\alpha$ is not free but is  a function of $(\vartheta, \zeta)$
and determined by
\ba
\label{nominalization}
1=  \int \exp\{ \alpha+ r(\bx, \vartheta)\} g(\bx, \zeta) d\bx.
\ea

The following three functions will be useful in our proof:
\bas
B_1(d, \bx, y)
&=&
\frac{\partial \log  h(d, \bx, y; \alpha_0, \vartheta_0, \eta_0, \zeta_0) }{
\partial \vartheta} \\
&=& (1-d) \{ \nabla_{\vartheta}\alpha(\vartheta_0)+\nabla_{\vartheta}
 r(\bx, \vartheta_0)\}+ d I_{e, -1} \nabla_{\xi} \log f(y|\bx, \xi_0), \\
B_2(d, \bx, y)
&=&
\frac{\partial \log  h(d, \bx, y; \alpha_0, \vartheta_0, \eta_0, \zeta_0) }{
\partial \eta} = \frac{D-\eta_0}{\eta_0(1-\eta_0)}, \\
B_3(d, \bx, y)
&=&
\frac{\partial \log  h(d, \bx, y; \alpha_0, \vartheta_0, \eta_0, \zeta_0) }{
\partial \zeta}=\nabla_{\zeta} \log g(\bx, \zeta_0),
\eas
where $I_{e, -1}$ is $I_e$ without the first row.

\subsection*{Semiparametric efficiency of $(\hat \theta, \hat \eta)$}

We have shown that
\bas
\hat \theta-\theta_0
&=&
n^{-1}\left(
\begin{array}{ccc}
V^{-1}-\frac{1}{\lambda_0(1-\lambda_0)}\bfe_1\bfe_1^\T\quad &\bfe_1\\
\end{array}
\right)u_n+o_p(n^{-1/2}), \\
\hat \eta - \eta_0
&=&
\frac{1}{n} \sum_{i=1}^n (d_i - \eta_0),
\eas
where
$u_n = (u_{n1}^\T, u_{n2})^\T$ with
\bas
u_{n1}
&=&
\sum_{i=1}^{n} \left[
(1-d_i - \pi_i ) \nabla_{\theta} t(\bx_i,\theta_0)
+ d_i I_e \nabla_{\xi}     \log \{f(y_i|\bx_i, \xi_0)\}\right],
\\
u_{n2}
&=& \frac{1}{\lambda_0(1-\lambda_0)}\sum_{i=1}^{n}\left(\lambda_0-\pi_i\right).
\eas
Therefore
\bas
\hat \theta-\theta_0
&=&
n^{-1} \{
  V^{-1}u_{n1} -\frac{1}{\lambda_0(1-\lambda_0)}\bfe_1\bfe_1^\T u_{n1} + \bfe_1 u_{n2} \} +o_p(n^{-1/2})\\
&=&
n^{-1} \{
  V^{-1}u_{n1} + \frac{1}{\eta_0(1-\eta_0)}\bfe_1\sum_{i=1}^{n}(d_i - \eta_0)  \} +o_p(n^{-1/2}) \\
&=&
n^{-1} \sum_{i=1}^n
\Big[
  V^{-1}(1-d_i - \pi_i ) \nabla_{\theta} t(\bx_i,\theta_0)
+ V^{-1}d_i I_e \nabla_{\xi}     \log \{f(y_i|\bx_i, \xi_0)\}\\
&&
   +\frac{1}{\eta_0(1-\eta_0)}\bfe_1 (d_i - \eta_0)  \}
  \Big] +o_p(n^{-1/2}).
\eas
Then the respective influence functions of $\hat \theta$ and $\hat \eta$
are
\bas
\varphi_{\theta}(D, \bX, Y)
&=&
  V^{-1}(1-D - \pi(\bX) ) \nabla_{\theta} t(\bX,\theta_0)
+ V^{-1}D I_e \nabla_{\xi}     \log \{f(Y|\bX, \xi_0)\} \\
&&
  + \frac{(D - \eta_0) }{\eta_0(1-\eta_0)}\bfe_1
\eas
and
$\varphi_{\eta}(D, \bX, Y) = D - \eta_0$.
We  prove only the semiparametric efficiency of $\hat \theta$;
the semiparametric efficiency  of $\hat \eta$ can be proved
in the same way  with less algebra.

Referring to the established theory for the semiparametric efficiency bound, for example Chapter 3 of
\citet{Bickel1993} and \citet{Newey1990},
we need to show only the following two results to establish the semiparametric efficiency of $\hat\theta$:
\begin{description}
\item[(a)] $\hat \theta$ is a regular estimator of $\theta_0$;
\item[(b)] there exists a parametric submodel with $h_{\psi}(d, \bx, \tilde y)$ the joint density of
$(D, \bX, \tilde Y)$
such that  the true model is $h_{0}(d, \bx, \tilde y)$ and
$$
\varphi_{\theta}(d, \bx, y)=  \frac{\partial \log h_{\psi}(d, \bx, \tilde y) }{\partial \psi}\bigg|_{\psi = 0}.
$$
\end{description}

\noindent{\bf Proof of (a)}

By Theorem 2 in \citet{Newey1990},  arguing $\hat \theta$ is a regular estimator of $\theta_0$
is equivalent to showing that
\ba
Z_1&\equiv & \e \{ \varphi_{\theta}(D, \bX, Y) B_1^\T(D, \bX, Y)\}
\nonumber\\
&=&
 \frac{\partial \theta}{\partial \vartheta^\T}\Big|_{(\theta_0, \eta_0, \zeta_0)}=( \nabla_{\vartheta}\alpha ,\; I_{d_{\vartheta}} )^\T , \label{eq-theta} \\
Z_2&\equiv & \e \{ \varphi_{\theta}(D, \bX, Y) B_2 (D, \bX, Y)\}
= \frac{\partial \theta }{\partial \eta}\Big|_{(\theta_0, \eta_0, \zeta_0)}  = 0,\label{eq-eta}\\
Z_3&\equiv & \e \{ \varphi_{\theta}(D, \bX, Y) B_3^\T(D, \bX, Y)\}
= \frac{\partial \theta }{\partial \zeta^\T}\Big|_{(\theta_0, \eta_0, \zeta_0)} =0,  \label{eq-zeta}
\ea
where throughout this section $\e$ takes expectation with respect to $  h(d, \bx,  y; \theta_0, \eta_0, \zeta_0) $.

{\it  (1)  Proof of Equality \eqref{eq-theta}}

Since
$
\e \{ D \nabla_{\xi}     \log  f(Y|\bX, \xi_0) |\bX\} = 0,
$
it follows that
\bas
Z_1&=& \e \{ \varphi_{\theta}(D, \bX, Y) B_1^\T(D, \bX, Y)\} \\
&=&
\e [ (1-D - \pi(\bX) ) V^{-1} \nabla_{\theta } t(\bX,\theta_0)   (1-D) \{ \nabla_{\vartheta}\alpha(\vartheta_0)
+\nabla_{\vartheta} r(\bX, \theta_0)\}^\T ] \\
&&
+ \e [(1-D)  r(\bX, \theta_0)
\frac{(D - \eta_0) }{\eta_0(1-\eta_0)}\bfe_1 \{ \nabla_{\vartheta}\alpha(\vartheta_0) +\nabla_{\vartheta} \}^\T
  ]  \\
&&
+
\e [  D   V^{-1}  I_{e } \nabla_{\xi}\{  \log  f(Y|\bX, \xi_0) \}^{\otimes 2} I_{e, -1}^\T  ]
  \eas
  \bas
&=&
\e [\pi(\bX)   (1 - \pi(\bX) )V^{-1} \nabla_{\theta } t(\bX,\theta_0)   \{ \nabla_{\vartheta}\alpha(\vartheta_0)
+\nabla_{\vartheta} r(\bX, \theta_0)\}^\T ] \\
&&
- \e [
\frac{(1-D)  \eta_0 }{\eta_0(1-\eta_0)}\bfe_1   \{ \nabla_{\vartheta}\alpha(\vartheta_0) +\nabla_{\vartheta} r(\bX, \theta_0)\}^\T   \}
  ]  \\
&&
+
\e [  D  I_{e} \nabla_{\xi}  \{  \log  f(Y|\bX, \xi_0) \}^{\otimes 2} I_{e, -1}^\T  V^{-1}  ]
\\
&=&
\bfe_1 \{ \nabla_{\vartheta}\alpha(\vartheta_0) \}^\T
+
\e [
\frac{ \pi(\bX) }{1-\eta_0}\bfe_1  \{ \nabla_{\vartheta}\alpha(\vartheta_0) +\nabla_{\vartheta} r(\bX, \theta_0)\}^\T  \}
  ]  +  I_{e}I_{e, -1}^\T,
\eas
where we have used the definition
\bas
V =   \e[  \{1-\pi(\bX)\} \pi(\bX)   \{ \nabla_\theta t(\bX, \theta  ) \}^{\otimes 2} ]
+
 \e[  D  I_e \{ \nabla_{\xi } f(Y|\bX,  \xi ) \}^{\otimes 2}I_e^\T ].
\eas

Taking derivative with respect to $\vartheta $ on both sides of \eqref{nominalization} gives
\bas
0 &=&  \int \{   \nabla_{\vartheta}\alpha(\vartheta_0) + \nabla_{\vartheta} r(\bx, \vartheta_0) \}
\exp\{ \alpha(\vartheta_0) + r(\bx, \vartheta_0)\} g(\bx, \zeta_0) d\bx.
\eas
This together with $g(\bx, \zeta_0) d\bx =  dF(\bx|D=1)$ leads to
\bas
&& \frac{1}{1-\eta_0}\bfe_1^\T
 \e [ \{ \nabla_{\vartheta}\alpha(\vartheta_0) +\nabla_{\vartheta} r(\bX, \theta_0)\}
  \pi(\bX)  \}
  ] \\
 &=&
 \frac{\eta_0}{1-\eta_0}\bfe_1^\T \int \{   \nabla_{\vartheta}\alpha(\vartheta_0) + \nabla_{\vartheta} r(\bx, \vartheta_0) \}
\exp\{ \alpha(\vartheta_0) + r(\bx, \vartheta_0)\} g(\bx, \zeta_0) d\bx \\
&=& 0.
 \eas
Therefore, we have
\bas
Z_1
&=&
 \nabla_{\vartheta}\alpha(\vartheta_0) e_1^\T
 + I_{e}I_{e, -1}^\T
 =
(\nabla_{\vartheta}\alpha, I_{d_{\vartheta}}^\T)^\T.
\eas
This proves  \eqref{eq-theta}.

{\it (2) Proof of Equality \eqref{eq-eta}}

Since $\e \varphi_{\theta}(D, \bX, Y) =0$, we have
\bas
Z_2&=&
\e\{B_2(D,\bX, Y) \varphi_{\theta}(D, \bX, Y) \}  \\
&=&
\frac{1}{\eta_0(1-\eta_0)} \e\{ D \varphi_{\theta}(D, \bX, Y) \}  \\
&=&
\frac{1}{\eta_0(1-\eta_0)} \e\{
 - D V^{-1}  \pi(\bX)   \nabla_{\theta} t(\bX,\theta_0)
 + D \frac{1}{\eta_0(1-\eta_0)}\bfe_1 (1 - \eta_0)  \}\\
    &&
+ D V^{-1}  I_e \nabla_{\xi}     \log \{f(Y|\bX, \xi_0)\}
 \}  \\
&=&
\frac{1}{\eta_0(1-\eta_0)} \e\{
 -  V^{-1} (1-\pi(\bX)) \pi(\bX)   \nabla_{\theta} t(\bX,\theta_0)
  +  \bfe_1
 \}  \\
 &=& 0,
\eas
where the last equality holds because
\bas
 \e\{ \pi(\bX) (1-\pi(\bX)) \nabla_{\theta}  t(\bX, \theta_0)\} = Ve_1.
\eas
This proves Equality \eqref{eq-eta}.

{\it (3)  Proof of Equality \eqref{eq-zeta}}

 Since
\bas
\e\{ \varphi_{\theta}(D, \bX, Y) |\bx \}
= \frac{1}{\eta_0(1-\eta_0)}\bfe_1 \{ 1 - \eta_0 - \pi(\bX)  \},
\eas
we have
\bas
Z_3
&=&
\e\{ \varphi_{\theta}(D, \bX, Y)  B_3^\T(D,\bX, Y) \} \\
&=&
\e[ \frac{1}{\eta_0(1-\eta_0)}\bfe_1  \{ 1 - \eta_0 - \pi(\bX)  \} \nabla_{\zeta^\T} \log g(\bX, \zeta_0)  ] \\
&=&
 -\frac{1}{\eta_0(1-\eta_0)} \e[  \bfe_1    \pi(\bX) \nabla_{\zeta^\T} \log g(\bX, \zeta_0)].
\eas

Taking derivative with respect to $\zeta$ on both sides of Eq \eqref{nominalization} gives
\bas
 0 &=&  \int \exp\{ \alpha_0 + r(\bx, \vartheta_0)\}\{\nabla_{\zeta} \log g(\bx, \zeta_0)\} g(\bx, \zeta_0) d\bx \\
   &=&  \int \exp\{ \alpha_0 + r(\bx, \vartheta_0)\}  \{\nabla_{\zeta} \log g(\bx, \zeta_0)\}  \frac{1-\pi(\bx)}{\eta_0} \pr(\bx) d\bx \\
  &=&  \int   \{\nabla_{\zeta} \log g(\bx, \zeta_0)\}  \frac{\pi(\bx)}{\eta_0(1-\eta_0)} \pr(\bx) d\bx \\
    &=& \e [  \{\nabla_{\zeta} \log g(\bX, \zeta_0)\}  \frac{\pi(\bX)}{\eta_0(1-\eta_0)} ],
\eas
which means
$
Z_3=0.
$
This proves Equality \eqref{eq-zeta} and also completes the proof of (a).

\noindent{\bf Proof of (b)}

Consider the following function
\bas
h_{\psi}(d, \bx, \tilde y)
&=&
\{ 1+\psi \varphi_{\theta}(d, \bx, y) \} \times
\{f(y|\bx, \xi_0) \eta_0  \}^d
 \\
  && \times  \{ \exp(\alpha_0+ r(\bx, \vartheta_0)) (1-\eta_0)  \}^{1-d} g(\bx, \zeta_0).
\eas
Suppose the support of $(\bX, Y)$ is compact, then
it can be verified that the function
\bas
\varphi_{\theta}(D, \bX, Y)
&=&
  V^{-1}(1-D - \pi(\bX) ) \nabla_{\theta} t(\bX,\theta_0)
 +\frac{1}{\eta_0(1-\eta_0)}\bfe_1 (D - \eta_0)  \\
    &&
+ V^{-1}D I_e \nabla_{\xi}     \log \{f(Y|\bX, \xi_0)\}
\eas
is bounded.
Because  $\e  \{ \varphi_{\theta}(D, \bX, Y) \}=0$ where $\e $
takes expectation with respect to
$ h(d, \bx, y; \alpha_0, \vartheta_0, \eta_0, \zeta_0)$,
the function
$h_{\psi}(d, \bx, \tilde y)$
is a density function
when $\psi$ is small enough.
When $\psi=0$, it reduces to
the true joint density function
$ h(d, \bx, y; \alpha_0, \vartheta_0, \eta_0, \zeta_0)$.
It is easy to check that
$h_{\psi}(d, \bx, \tilde y)$ with small enough $\psi$
is a parametric submodel
and
\bas
\nabla_{\psi}  h_{\psi}(d, \bx, \tilde y) \Big|_{\psi=0} =  \varphi_{\theta}(d, \bx, y).
\eas
This proves (b), and hence proves the semiparametric efficiency of $\hat \theta$.

\subsection*{Semiparametric efficiency of $\hat \mu$}

The population mean can be expressed as
\bas
\mu
&=&
\int_y \int_{\bx} y \pr(y|\bx, D=1) \pr(\bx|D=1) \pr(D=1) d\bx dy \\
&& +
 \int_y \int_{\bx} y \pr(y|\bx, D=0) \pr(\bx|D=0) \pr(D=0) d\bx dy  \\
&=&
 \int_y \int_{\bx} y \pr(y|\bx, D=1) \pr(\bx|D=1) \eta d\bx dy \\
 && +
\int_y \int_{\bx} y  \exp(\alpha + \bx^\T\beta  +\gamma y)  \pr(y|\bx, D=1) \pr(\bx|D=1) (1-\eta) d\bx dy \\
&=&
\int_{\bx} \left[ \int_y  y \{ \eta +(1-\eta)\exp(\alpha + \bx^\T\beta  +\gamma y) \} f(y|\bx, \xi)dy\right] dF(\bx|D=1).
\eas
The proposed mean estimator is
\bas
\hat \mu  &=&
  \frac{1}{n} \sum_{i=1}^n \frac{  \int_y  y \{ \hat \eta +(1-\hat \eta)
       \exp(\hat \alpha + \bx_i^\T\hat \beta  +\hat \gamma y) \} f(y|\bx_i, \hat \xi)dy  }{
  \hat \eta + (1-\hat \eta) \exp\{ \hat \alpha +\bx_i^\T\hat \beta+c(\bx_i,\hat \gamma,\hat \xi)\} } \\
&=& n^{-1} \sum_{i=1}^n K(\bx_i;\hat \theta, \hat \eta),
\eas
where
\bas
K(\bx;\theta,\eta)&=&
 \frac{
\int  y \{  \eta + (1-\eta)\exp(  \alpha + \bx^\T \beta  + \gamma y) \} f(y|\bx,   \xi)dy  }{
\eta + (1-\eta) \exp\{  \alpha +\bx^\T  \beta+c(\bx, \gamma,  \xi)\} }.
\eas

Recall that $A=\e\left\{ \nabla_{\theta}K(\bX;\theta_0,\eta_0)\right\}$,
$\pi_i=\pi(\bx_i)$ and
\(
I_e^\T  = (0_{d_{\xi}\times (2+d_{\beta})}, I_{d_\xi\times d_{\xi}}).
\)
We have shown in the proof of Theorem 2 that
$$
\hat\mu-\mu
=\frac{1}{n}\sum_{i=1}^n\{K(\bx_i;\theta_0,\eta_0)-\mu\}+n^{-1}A^\T V^{-1} u_{n1}+o_p(n^{-1/2}),
$$
where
$
u_{n1}
=
\sum_{i=1}^{n} \left[
(1-d_i - \pi_i ) \nabla_{\theta} t(\bx_i,\theta_0)
+ d_i I_e \nabla_{\xi}     \log \{f(y_i|\bx_i, \xi_0)\}\right].
$
Equivalently
\bas
\hat\mu-\mu
&=&
\frac{1}{n}\sum_{i=1}^n\{ K(\bx_i;\theta_0,\eta_0)-\mu_0 +
(1-d_i - \pi_i )  A^\T V^{-1}
\nabla_{\theta} t(\bx_i,\theta_0) \\
&&
+ d_i A^\T V^{-1}  I_e \nabla_{\xi}     \log  f(y_i|\bx_i, \xi_0)\} +o_p(n^{-1/2}),
\eas
which implies that the influence function of $\hat \mu$ is
\bas
\varphi_{\mu}(D, \bX, Y)
&=&   K(\bX;\theta_0,\eta_0)-\mu_0 +
\{ 1-D - \pi(\bX) \}  A^\T V^{-1}
\nabla_{\theta} t(\bX,\theta_0) \\
&&
+ D  A^\T V^{-1}  I_e \nabla_{\xi}     \log  f(Y|\bX, \xi_0).
\eas

Similar to the proof of the semiparametric efficiency of $\hat \theta$,
we need to show only the following two results to establish the semiparametric efficiency of $\hat\mu$:
\begin{description}
\item[(a1)] $\hat \mu$ is a regular estimator of $\mu_0$;
\item[(b1)] there exists a parametric submodel with $h^*_{\psi}(d, \bx, \tilde y)$ the joint density of
$(D, \bX, \tilde Y)$
such that  the true model is $h^*_{0}(d, \bx, \tilde y)$ and
$$
\varphi_{\mu}(d, \bx, y)=  \frac{\partial \log h^*_{\psi}(d, \bx, \tilde y) }{\partial \psi}\bigg|_{\psi = 0}.
$$
\end{description}

\noindent{\bf Proof of (a1)}

Under the submodel $g(\bx, \zeta)$ for $\pr(\bX=\bx|D=1)$,
we can write $\mu$ as
\bas
\mu
= \mu(\theta, \eta, \zeta)
 \equiv
\int_{\bx} \left[ \int_y  y \{ \eta +(1-\eta)\exp(\alpha + \bx^\T\beta  +\gamma y) \} f(y|\bx, \xi)dy\right]
g(\bx, \zeta) d\bx.
\eas
Define  $w(\bx, y) = (\bx^\T, y, \nabla_{\xi^\T} \log f(y|\bx, \xi_0))^\T$.
The partial derivative of $ \mu$
is
\bas
\nabla_{\vartheta} \mu (\theta_0, \eta_0)
&=&
\int_{\bx}   \int_y  y \{\eta_0+ (1-\eta_0) \exp(\alpha_0 + \bx^\T\beta_0  +\gamma_0 y) \}  \\
&&
\times
\{  \nabla_{\vartheta} \alpha(\vartheta_0) + w(\bx, y) \}  f(y|\bx, \xi_0)dy  g(\bx, \zeta) d\bx.
\eas

By Theorem 2 in \citet{Newey1990},  arguing $\hat \mu$ is a regular estimator of $\mu_0$
is equivalent to showing that
\ba
C_1&\equiv & \e \{ \varphi_{\mu}(D, \bX, Y) B_1(D, \bX, Y)\}
= \frac{\partial \mu (\theta_0, \eta_0, \zeta_0)}{\partial \vartheta}, \label{eq-theta1} \\
C_2&\equiv & \e \{ \varphi_{\mu}(D, \bX, Y) B_2(D, \bX, Y)\}
= \frac{\partial \mu (\theta_0, \eta_0, \zeta_0)}{\partial \eta}, \label{eq-eta1}\\
C_3&\equiv & \e \{ \varphi_{\mu}(D, \bX, Y) B_3(D, \bX, Y)\}
= \frac{\partial \mu (\theta_0, \eta_0, \zeta_0)}{\partial \zeta},  \label{eq-zeta1}
\ea
where   $\e$ takes expectation with respect to $  h(d, \bx,  y; \theta_0, \eta_0, \zeta_0) $.
Keep in mind that $\alpha$ is a function of $\vartheta$ and $\zeta$.

{\it (1) Proof of Equality \eqref{eq-theta1}}

Since
\(
\e \{ D \nabla_{\xi}     \log  f(Y|\bX, \xi_0) |\bX\} = 0,
\)
it follows that
\bas
C_1&=& \e \{ \varphi_{\mu}(D, \bX, Y) B_1(D, \bX, Y)\} \\
&=&
\e [(1-D) \nabla_{\theta} \{ \nabla_{\vartheta}\alpha(\vartheta_0)
+\nabla_{\vartheta} r(\bX, \theta_0)\} \{ K(\bX;\theta_0,\eta_0)-\mu_0  \}] \\
&&
+ \e [(1-D)\{ 1-D - \pi(\bX) \} \{ \nabla_{\vartheta}\alpha(\vartheta_0) +\nabla_{\vartheta} r(\bX, \theta_0)\}
 \nabla_{\theta^\T} t(\bX,\theta_0)  V^{-1} A
  ]
  \eas\bas
&&
+
\e [  D \nabla_{\xi}  I_{e, -1} \{  \log  f(Y|\bX, \xi_0) \}^{\otimes 2} I_e^\T  V^{-1}    A ]
\\
&=&
\e [ \pi(\bX )   \{ \nabla_{\vartheta}\alpha(\vartheta_0) +\nabla_{\vartheta} r(\bX, \theta_0)\} \{ K(\bX;\theta_0,\eta_0)-\mu_0  \}] \\
&&
+ \e [\pi(\bX) \{ 1 - \pi(\bX) \}  \{\nabla_{\vartheta} t(\bX, \theta_0)\}^{\otimes2 } V^{-1} A
   ] \\
&&   +
\e [  D \nabla_{\xi} I_{e, -1}  \{  \log  f(Y|\bX, \xi_0) \}^{\otimes 2} I_e^\T  V^{-1}    A ]  \\
&=&
\e [ \pi(\bX ) \{ \nabla_{\vartheta}\alpha(\vartheta_0) +\nabla_{\vartheta} r(\bX, \theta_0)\}
 \{ K(\bX;\theta_0,\eta_0)-\mu_0  \}]+   A_{-1},
\eas
where
$A_{-1}$ is $A$ without its first component  and
we have used the definition of $V$.

Because
\bas
 && \nabla_{\vartheta} K(\bx;\theta_0,\lambda_0)   \\
&=&
 \frac{
\int  y \{ \nabla_{\vartheta}\alpha(\vartheta_0) + w(\bx, y) \}
\lambda_0 \exp(  \alpha_0  + \bx^\T \beta_0  + \gamma_0 y)   f(y|\bx,   \xi_0)dy  }{
(1-\lambda_0)+ \lambda_0 \exp\{  t(\bx, \theta_0)\} }     \\
&&
-
K(\bx;\theta_0,\lambda_0)
 \frac{\lambda_0 \exp\{  t(\bx, \theta_0)\}  }{
  (1-\lambda_0)+ \lambda_0 \exp\{  t(\bx, \theta_0)\}  }  \{\nabla_{\vartheta}\alpha(\vartheta_0)
  + \nabla_{\vartheta} r(\bx, \theta_0)\} \\
  &=&
  \frac{1-\eta_0}{\eta_0} \{1- \pi(\bx)\}\cdot
\int  y \{ \nabla_{\vartheta}\alpha(\vartheta_0) + w(\bx, y) \}
  \exp(  \alpha_0  + \bx^\T \beta_0  + \gamma_0 y)   f(y|\bx,   \xi_0)dy       \\
&&
-
K(\bx;\theta_0,\lambda_0)
\pi(\bx) \{\nabla_{\vartheta}\alpha(\vartheta_0)
  + \nabla_{\vartheta} r(\bx, \theta_0)\},
\eas
we have
\bas
A_{-1}
&=&
\e\{ \nabla_{\vartheta} K(\bx;\theta_0,\lambda_0) \}  \\
&=&
\e[ \frac{1-\eta_0}{\eta_0}\{1- \pi(\bx)\}
\int  y  \{ \nabla_{\vartheta}\alpha(\vartheta_0) + w(\bx, y) \} \exp(  \alpha_0 + \bx^\T \beta_0  + \gamma_0 y)   f(y|\bx,   \xi_0)dy] \\
&&
-
\e [ K(\bx;\theta_0,\lambda_0) \pi(\bx) \{\nabla_{\vartheta}\alpha(\vartheta_0)
  + \nabla_{\vartheta} r(\bx, \theta_0)\} ]
\eas
\bas
&=&
\int (1-\eta_0)
\int  y  \{ \nabla_{\vartheta}\alpha(\vartheta_0) + w(\bx, y) \}  \exp(  \alpha_0 + \bx^\T \beta_0  + \gamma_0 y)   f(y|\bx,   \xi_0)dy] dF(\bx|D=1)\\
&&
-
\int  K(\bx;\theta_0,\lambda_0) (1-\eta_0)\exp\{  t(\bx, \theta_0)\}  \{\nabla_{\vartheta}\alpha(\vartheta_0)
  + \nabla_{\vartheta} r(\bx, \theta_0)\}  dF(\bx|D=1),
\eas
where we have used $ t(\bx, \theta)= \alpha + r(\bx, \vartheta)$ and
\bas
\pr(\bx) = \frac{\eta_0}{1-\pi(\bx)} \pr(\bx|D=1).
\eas

It follows that
\bas
C_1 &=&
\e [ \pi(\bX )  \{ \nabla_{\vartheta}\alpha(\vartheta_0) +\nabla_{\vartheta} r(\bX, \vartheta_0)\}
\{ K(\bX;\theta_0,\eta_0)-\mu_0  \}]+   A_{-1}  \\
&=&
\int  (1-\eta_0)\exp\{  t(\bx, \theta_0)\}     \{ \nabla_{\beta}\alpha(\vartheta_0) + \nabla_{\vartheta} r(\bx, \vartheta_0\} \{  K(\bx;\theta_0,\lambda_0) - \mu_0\}  dF(\bx|D=1)
  +   \\
&&
+
\int (1-\eta_0)
\int  y  \{ \nabla_{\vartheta}\alpha(\vartheta_0) + w(\bx, y) \}  \exp(  \alpha_0 + \bx^\T \beta_0  + \gamma_0 y)   f(y|\bx,   \xi_0)dy] dF(\bx|D=1)\\
&&
-
\int  K(\bx;\theta_0,\lambda_0) (1-\eta_0)\exp\{  t(\bx, \theta_0)\}  \{\nabla_{\vartheta}\alpha(\vartheta_0)
  + \nabla_{\vartheta} r(\bx, \theta_0)\}  dF(\bx|D=1) \\
&=&
- \mu_0 \int  (1-\eta_0)\exp\{  t(\bx, \theta_0)\}     \{ \nabla_{\beta}\alpha(\vartheta_0) + \nabla_{\vartheta} r(\bx, \vartheta_0\} dF(\bx|D=1)
  +   \\
&&
+
\int (1-\eta_0)
\int  y  \{ \nabla_{\vartheta}\alpha(\vartheta_0) + w(\bx, y) \}  \exp(  \alpha_0 + \bx^\T \beta_0  + \gamma_0 y)   f(y|\bx,   \xi_0)dy] dF(\bx|D=1).
\eas
Taking derivative with respect to $\vartheta $ on both sides of \eqref{nominalization} gives
\bas
0 &=&  \int \{   \nabla_{\vartheta}\alpha(\vartheta_0) + \nabla_{\vartheta} r(\bx, \vartheta_0) \}
\exp\{ \alpha(\vartheta_0) + r(\bx, \vartheta_0)\} g(\bx, \zeta_0) d\bx.
\eas
Since $g(\bx, \zeta_0) d\bx=  \pr(\bx|D=1) d\bx =  dF(\bx|D=1)$,
it follows that
\bas
C_1 &=&
 \int (1-\eta_0)
\int  y  \{ \nabla_{\vartheta}\alpha(\vartheta_0) + w(\bx, y) \}  \exp(  \alpha_0 + \bx^\T \beta_0  + \gamma_0 y)   f(y|\bx,   \xi_0)dy] dF(\bx|D=1),
\eas
which is exactly $\nabla_{\vartheta} \mu (\theta_0, \eta_0, \zeta_0)$.

{\it (2) Proof of Equality \eqref{eq-eta1}}

Since $\e \varphi_{\mu}(D, \bX, Y) =0$, we have
\bas
C_2&=&
\e\{B_2(D,\bX, Y) \varphi_{\mu}(D, \bX, Y) \}  \\
&=&
\frac{1}{\eta_0(1-\eta_0)} \e\{ D \varphi_{\mu}(D, \bX, Y) \}  \\
&=&
\frac{1}{\eta_0(1-\eta_0)} \e\{ D
 K(\bX;\theta_0,\eta_0)-\mu_0 D
   - \pi(\bX) D  A^\T V^{-1}
\nabla_{\theta} t(\bX,\theta_0) \\
&&
+ D  A^\T V^{-1}  I_e \nabla_{\xi}     \log  f(Y|\bX, \xi_0)
 \}  \\
&=&
\frac{1}{\eta_0(1-\eta_0)} \e\{ (1-\pi(\bX))
 K(\bX;\theta_0,\eta_0)-\mu_0 (1-\pi(\bX)) \\
&&
   - \pi(\bX) (1-\pi(\bX)) A^\T V^{-1}
\nabla_{\theta} t(\bX,\theta_0)
 \}  \\
 &=&
\frac{1}{\eta_0(1-\eta_0)}
[\e\{ (1-\pi(\bX))  K(\bX;\theta_0,\eta_0) \}
-\mu_0 \eta_0     -  A^\T \bfe_1
 ],
\eas
where the last equality holds because
$
V\bfe_1
= \e\{ \pi(\bX) (1-\pi(\bX)) \nabla_{\theta}  t(\bX, \theta_0)\}.
$

Meanwhile because
\bas
 A^\T \bfe_1
&=&
\e\{ \nabla_{\alpha} K(\bx;\theta_0,\lambda_0) \}  \\
&=&
\e[ \frac{1-\eta_0}{\eta_0}\{1- \pi(\bX)\}
\int  y   \exp(  \alpha_0 + \bX^\T \beta_0  + \gamma_0 y)   f(y|\bX,   \xi_0)dy]
-
\e \{ K(\bx;\theta_0,\lambda_0) \pi(\bx) \},
\eas
we further have
\bas
C_2 \eta_0(1-\eta_0)
&=&
 \e\{ (1-\pi(\bX))  K(\bX;\theta_0,\eta_0) \}
-\mu_0 \eta_0  \\
&&    - \e[ \frac{1-\eta_0}{\eta_0}\{1- \pi(\bX)\}
\int  y   \exp(  \alpha_0 + \bX^\T \beta_0  + \gamma_0 y)   f(y|\bX,   \xi_0)dy]  \\
&&
+
\e [  K(\bX;\theta_0,\eta_0) \pi(\bX) ]
\eas
\bas
&=&
 \e\{   K(\bX;\theta_0,\eta_0) \}
-\mu_0 \eta_0  \\
&&    -  \e[ \frac{1-\eta_0}{\eta_0}\{1- \pi(\bX)\}
\int  y   \exp(  \alpha_0 + \bX^\T \beta_0  + \gamma_0 y)   f(y|\bX,   \xi_0)dy] \\
 &=&
\mu_0(1 - \eta_0)    -  (1 - \eta_0)
\int
\int  y   \exp(  \alpha_0 + \bX^\T \beta_0  + \gamma_0 y)   f(y|\bX,   \xi_0)dy dF(\bx|D=1).
\eas
Using the definition of $\mu_0$, we have
\bas
&& C_2 (1 - \eta_0) \eta_0  \\
 &=&
 (1 - \eta_0)  \int_{\bx} \left[ \int_y  y \{ \eta +(1-\eta)\exp(\alpha + \bx^\T\beta  +\gamma y) \} f(y|\bx, \xi)dy\right] dF(\bx|D=1)  \\
&&
-  (1 - \eta_0)
\int
\int  y   \exp(  \alpha_0 + \bX^\T \beta_0  + \gamma_0 y)   f(y|\bX,   \xi_0)dy \pr(\bx|D=1) d\bx  \\
 &=&
 (1 - \eta_0)  \int_{\bx} \left[ \int_y  y \{ \eta_0   - \eta_0 \exp(\alpha_0 + \bx^\T\beta_0  +\gamma_0 y) \}
 f(y|\bx, \xi)dy\right] dF(\bx|D=1)   \\
  &=&
 (1 - \eta_0) \eta_0 \int_{\bx} \left[ \int_y  y \{ 1   -  \exp(\alpha_0 + \bx^\T\beta_0  +\gamma_0 y)  \}
 f(y|\bx, \xi)dy\right] dF(\bx|D=1).
\eas

Since
\bas
\nabla_{\eta} \mu
&=&
\int_{\bx} \left[ \int_y  y \{ 1  - \exp(\alpha + \bx^\T\beta  +\gamma y) \} f(y|\bx, \xi)dy\right] dF(\bx|D=1),
\eas
we arrive at
\bas
C_2 (1 - \eta_0) \eta_0
 &=&
 (1 - \eta_0) \eta_0 \nabla_{\eta} \mu(\theta_0, \eta_0, \zeta_0)
\Longleftrightarrow
 C_2=\nabla_{\eta} \mu(\theta_0, \eta_0, \zeta_0).
\eas
This proves Equality \eqref{eq-eta1}.

{\it (3) Proof of Equality \eqref{eq-zeta1}}

Since
$
\e\{ \varphi_{\mu}(D, \bX, Y) |D=1 \}
= K(\bX;\theta_0,\eta_0)-\mu_0,
$
we have
\bas
C_3
&=&
\e\{ B_3(D,\bX, Y) \varphi_{\mu}(D, \bX, Y) \} \\
&=&
\e[ \nabla_{\zeta} \log g(\bX, \zeta_0)\{  K(\bX;\theta_0,\eta_0)-\mu_0 \} ] \\
&=&
\e[ \nabla_{\zeta} \log g(\bX, \zeta_0)  K(\bX;\theta_0,\eta_0) ].
\eas

Note that
\bas
\mu
=
\e\{  K(\bX;\theta_0,\eta_0) \}
=
\int_{\bx} K(\bx, \theta_0, \eta_0) [  \eta_0 + (1-\eta_0) \exp\{t(\bx, \theta_0)\}  ]  dF(\bx|D=1),
\eas
which implies
\bas
\nabla_{\zeta}
\mu
&=&
\int_{\bx} K(\bx, \theta_0, \lambda_0) [  \eta_0 + (1-\eta_0) \exp\{t(\bx, \theta_0)\}  ]
\{ \nabla_{\zeta}\log g(\bx, \zeta_0) \} dF(\bx|D=1) \\
&=&
\int_{\bx} K(\bx, \theta_0, \lambda_0)
\{ \nabla_{\zeta}\log g(\bx, \zeta_0) \} \pr(\bx) d\bx \\
&=& \e[ \nabla_{\zeta} \log g(\bX, \zeta_0)  K(\bX;\theta_0,\eta_0) ] \\
&=& C_3.
\eas
This proves Equality \eqref{eq-zeta1} and also completes the proof of (a1).

\noindent{\bf Proof of (b1)}

Consider the following function
\bas
h_{\psi}^*(d, \bx, \tilde y)
&=&
\{ 1+\psi \varphi_{\mu}(d, \bx, y) \} \times
\{f(y|\bx, \xi_0) \eta_0  \}^d
  \{ \exp(\alpha_0+ r(\bx, \vartheta_0)) (1-\eta_0)  \}^{1-d} \\
  &&  \times g(\bx, \zeta_0).
\eas
Suppose the support of $(\bX, Y)$ is compact, then
it can be verified that the function
 \bas
\varphi_{\mu}(D, \bX, Y)
&=&    K(\bX;\theta_0,\eta_0)-\mu_0 +
\{ 1-D - \pi(\bX) \}  A^\T V^{-1}
\nabla_{\theta} t(\bX,\theta_0) \\
&&
+ D  A^\T V^{-1}  I_e \nabla_{\xi}     \log  f(Y|\bX, \xi_0)
\eas
is bounded.
Because  $\e  \{ \varphi_{\mu}(D, \bX, Y) \} = 0$ where $\e $
takes expectation with respect to
$ h(d, \bx, y; \alpha_0, \vartheta_0, \eta_0, \zeta_0)$,
the function
$h_{\psi}^*(d, \bx, \tilde y) $
is a density function
when $\psi$ is small enough.
When $\psi=0$, it reduces to
the true joint density function
$ h(d, \bx, y; \alpha_0, \vartheta_0, \eta_0, \zeta_0)$.
It is easy to check that
$h_{\psi}^*(d, \bx, \tilde y)$ with small enough $\psi$
is a parametric submodel
and
\bas
\nabla_{\psi}  h_{\psi}^*(d, \bx, \tilde y) \Big|_{\psi=0} =  \varphi_{\mu}(d, \bx, y).
\eas
This proves (b1), and hence the semiparametric efficiency of $\hat \mu$.

\end{document}